\documentclass[authoryear,5p]{elsarticle}
\usepackage{amsmath}
\usepackage{eurosym}
\usepackage{color}
\usepackage{hyperref}
\numberwithin{equation}{section}

\begin{document}
\author[hevs,epfl]{Laurent Pagnier}
\ead{laurent.pagnier@epfl.ch}
\author[hevs]{Philippe Jacquod}
\ead{philippe.jacquod@hevs.ch}
\address[hevs]{School of Engineering, University of Applied Sciences of Western Switzerland HES-SO CH-1951 Sion, Switzerland}
\address[epfl]{Institute of Theoretical Physics, EPF Lausanne, CH-1015 Lausanne, Switzerland}
\title{How fast can one overcome the paradox of the energy transition? A physico-economic model for the European power grid}
\begin{abstract}
The paradox of the energy transition is that the low marginal costs of new renewable energy sources (RES) drag electricity prices down and discourage investments in flexible productions that are needed to compensate for the lack of dispatchability of the new RES.  The energy transition thus discourages the investments that are required for its own harmonious expansion. To investigate how this paradox can be overcome, we argue that, under certain assumptions, future electricity prices are rather accurately modeled from the residual load obtained by subtracting non-flexible productions from the load. Armed with the resulting economic indicator, we investigate future revenues for European power plants with various degree of flexibility. We find that, if neither carbon taxes nor fuel prices change, flexible productions would be financially rewarded better and sooner if the energy transition proceeds faster but at more or less constant total production, i.e. by reducing the production of thermal power plants at the same rate as the RES production increases. Less flexible productions, on the other hand, would see their revenue grow more moderately. Our results indicate that a faster energy transition with a quicker withdrawal of thermal power plants would reward flexible productions faster.
\end{abstract}
\maketitle

\section{Introduction}

\subsection{The paradox of the energy transition}
The goal of the energy transition is to meet energy demand from human activities in a sustainable way. In the electricity sector, the transition currently increases
the penetration of productions from new renewable energy sources (RES), in particular solar photovoltaic panels (PV) and wind turbines (WT). 
These RES differ from the traditional productions they substitute for, in at least two very significant ways. First, they 
lack dispatchability and have little mechanical inertia, second, they have very low marginal production costs. Their lack of dispatchability and mechanical inertia
requires additional flexible productions and possibly electrical energy storage (EES) to ensure the stability of the power grid as well as the balance of demand and supply at all times. 
Therefore increasing penetrations of new RES should be accompanied by
significant investments in new facilities with rather long payback periods.
However, the new RES's low marginal cost brings spot electricity prices and thus beneficiary margins of electric power companies down, while
further extending the payback period for investments in new facilities. The energy transition is therefore confronted with the paradox that it creates economic 
conditions which, at least temporarily, strongly discourage the infrastructural investments it needs to progress further. 
Because of that, a number of hydroelectric plant projects are currently frozen in Europe. To plan the next steps in the energy transition, to evaluate 
and anticipate the investments needed for its safe, steady progress, 
it is therefore important to get a relatively good quantitative estimate of future electricity prices. 
The key issue is whether production flexibility will soon be rewarded well enough
that it will motivate investments in fast dispatchable power plants and EES at a level consistent with the rate at which RES penetration increases.

The traditional way to address such questions is to construct economic models for electricity production and consumption. 
Those models are standardly based on a number of assumptions on 
general economic conditions, demographic evolution, costs of different fuels, maintenance and production costs, 
amount of taxes and subsidies on energy production and so forth. Once all these ingredients
are fixed, both the electricity demand and the marginal cost of different productions can be estimated, which determine the 
market price of electricity, from which one finally
computes expected future revenues. From these revenues, investment decisions can ultimately be made.
The accuracy of this procedure relies on the accuracy of each of the assumptions on which it is based. Unfortunately, the latter are to a high degree arbitrary -- 
economic growth rates, unemployment rates (indicative of the volume of industrial activities), taxation amounts, fuel costs (in particular natural gas prices), carbon taxes and so forth cannot be predicted accurately
on time scales of decades, corresponding to typical payback times for investments in the energy sector. In this manuscript we take a deliberately different approach, 
using as few working hypotheses as possible. {\it Eurotranselec}, our model to be presented below, is mostly based
on physical conditions extracted from the size and production types of national 
power plant fleets as well as the electricity demand. We argue that it introduces a reliable though quite simple
procedure to evaluate future electricity prices in the not too far future, 
given scenarios for the energy transition and the resulting evolution of power plant fleets.
We use it to investigate revenues of electric power plants in a time window until 2020. 

\begin{figure}
\center
\includegraphics[width=\columnwidth]{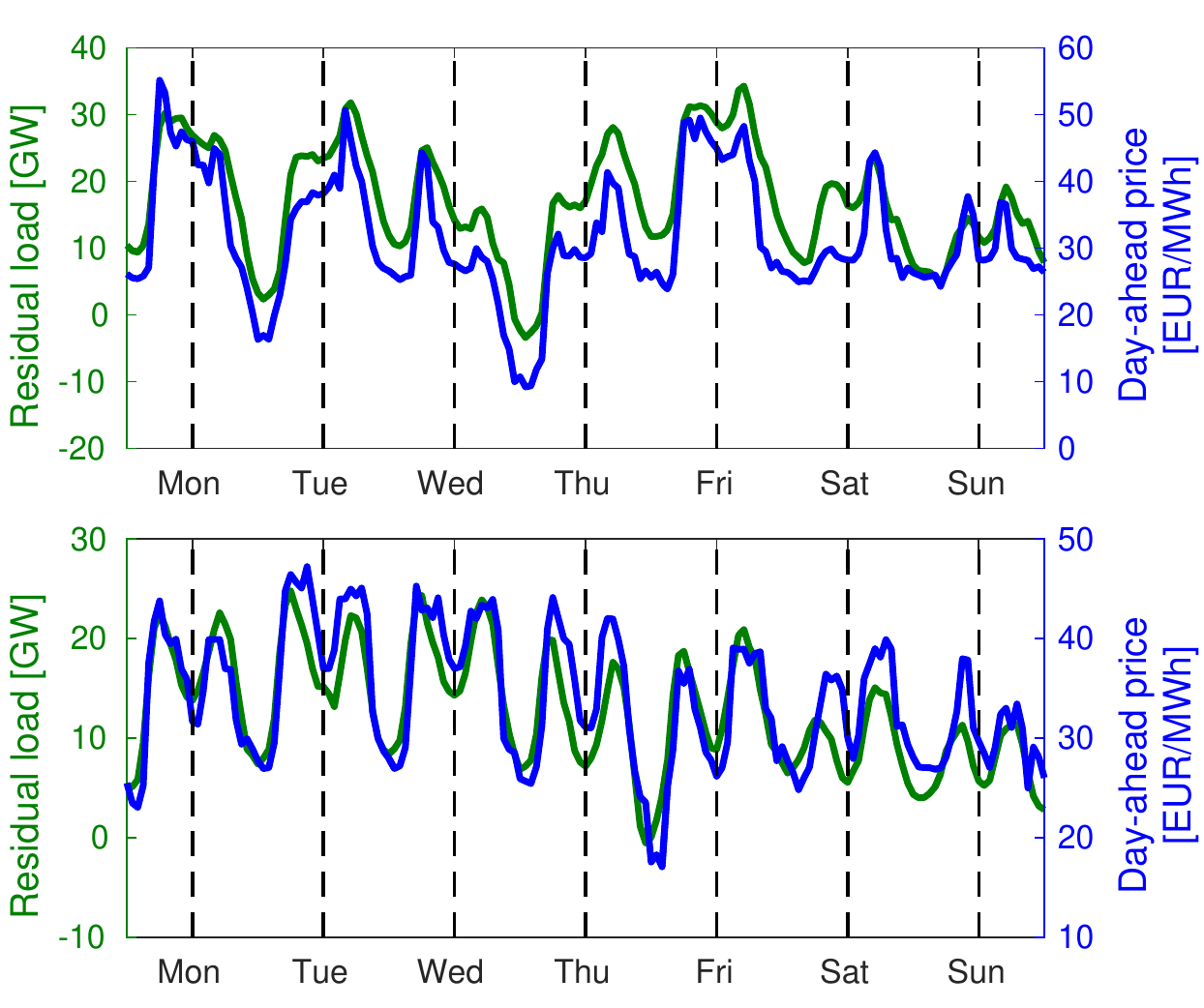}
\caption{German residual load (green) and day-ahead electricity price (blue) for a winter (top) and summer (bottom) week in 2015 [data taken from \cite{entsoe2015transparency}]. Vertical dashed lines indicate noon time.}\label{def_rl}
\end{figure}

\subsection{The residual load and electricity prices}
Our starting observation is that electricity prices reflect the law of supply and demand. Accordingly they exhibit some degree of correlation with 
what flexible sources must generate to sustain load -- the larger the difference between demand and non-dispatchable supply, the higher the electricity price. 
As a matter of fact, spot market prices are usually higher at times of larger imbalance between demand and non-dispatchable supply, when the imbalance
leans strongly on the demand side. The corresponding missing amount of power is quantified by the {\it residual load}, which is defined
as the difference between consumption and the sum of all non-flexible productions. With the residual load, the new RES 
are accounted for on the demand side and not as a production - they are seen as reducing the demand for the rest of the market.
This is justified by their almost vanishing marginal production cost.

While some degree of correlation between the residual load and spot market prices
is expected, a strong correlation between them
has been reported by \cite{roon2010modeling} for the special case of Germany in 2007--2009, with a coefficient of determination $R^2 \in [0.5,0.77]$.
Fig.~\ref{def_rl} confirms that a high degree of correlation between residual load and day-ahead market price prevails in Germany in 2015, which is quantified by a high value
of Pearson's correlation coefficient [see Eq.~\eqref{eq:coef} below] $r = 0.89$.\footnote{Assuming a linear relation between day-ahead prices and residual load, as we do in Eq.~\eqref{eq:lin_correl},  
with coefficients determined by a least square fitting, $R^2$ is the square of Pearson's correlation coefficient. Our finding of $r=0.89$ for Germany in 2015 then correponds to $R^2=0.8$, even higher than the
largest value reported by \cite{roon2010modeling}.} Below we investigate the correlation between day-ahead prices and residual load further, for
different European countries. We find that it  always corresponds to a large, positive correlation coefficient, $r>0.5$, 
and furthermore that $r$ is generally higher in countries with larger penetration of RES (this is shown in Fig.~\ref{corr} and will be discussed below). 
The energy transition will keep increasing the penetration of RES, 
it is therefore reasonable to expect that $r$ will increase in the future. If this is confirmed, the residual load will reflect
the day-ahead price better and better. 
Day-ahead transactions represent a significant percentage of all electricity transactions (see Table~\ref{da_table} below), and  
their share of the total load is expected to increase with the end of long-term 
contracts in the liberalized European electricity market. Putting all this together, we propose to introduce a synthetic electricity price $p_{\mathrm{da}}(t)$ as
a two-parameter, linear regression of the day-ahead price based on the residual load $R(t)$,
\begin{equation}\label{eq:lin_correl}
p_{\mathrm{da}}(t)=\Delta p_{\mathrm{da}} \, R(t)+p_{\mathrm{da}0} \, .
\end{equation}
In this manuscript, the two parameters $\Delta p_{\mathrm{da}}$ and $p_{\mathrm{da}0}$ are determined country by country from a least square fit of 2015 day-ahead data with Eq~\eqref{eq:lin_correl}. They are assumed to remain constant in the future, because we
consider a relatively small time window, until 2020.
However different scenarios can be considered, where $\Delta p_{\mathrm{da}}$ and $p_{\mathrm{da}0}$
evolve in time, for instance because economic conditions (fuel prices, subsidies and taxes) vary. Below we discuss in particular 
how $\Delta p_{\mathrm{da}}$ and $p_{\mathrm{da}0}$ qualitatively vary with varying carbon taxes and natural gas prices. 

We think that $p_{\mathrm{da}}(t)$ is a reliable day-ahead electricity price because it reproduces qualitatively and even almost always quantitatively historical time series for
the true spot market day-ahead price of electricity in all European countries we focus on. Not caught by our approach are extreme events, for instance corresponding to 
record low (high) RES productions with simultaneous record high (low) demand, giving unusually high price maxima (unusually low price minima). 

Armed with $p_{\mathrm{da}}(t)$ we finally investigate the parallel evolution of the 
energy transition and the electricity prices in Germany and Spain and compute expected future revenues for various types of power plants, focusing on  
pumped-storage and conventional dam hydroelectric plants. 
Under our assumptions, that neither consumption, nor gas prices will change significantly in the foreseeable future, our investigations indicate that revenues will bounce back faster and higher if RES production increases faster and if this transition is accompanied by a simultaneous phasing-out of superfluous thermal productions.

\subsection{Literature review}
Studies of the impact of increased penetration of new RES on electricity prices abound. Many of them investigated historical data vs.
the penetration of new RES to empirically express electricity prices as a function of green in-feed~\citep{clo2015merit,paraschiv2014impact}. 
In the liberalized European market, electricity prices are determined by a merit order supply curve where production capacities are ordered according to their
marginal costs. The effect of such merit order on historical electricity prices under increased penetration of new RES has been investigated by \cite{cludius2014merit}, who 
extrapolated their findings to evaluate future revenues of PV and WT. Going further, 
a number of studies investigated electricity markets where prices are determined by simulated merit orders with marginal costs as inputs~\citep{sensfuss2008merit,haas2013looming,auer2016integrating}, 
which often rely on self-consistent optimizations. As interesting as these works are, they are based on heavy algorithms as well as many assumptions (for instance 
future fuel prices) to build the merit order. \cite{schlachtberger2016backup} find that more flexible production is required as the penetration of new RES increases and that flexible sources become essential when the penetration of new RES reaches 50\%.

By definition, the residual load gives indications on periods of surplus or deficit of production of new RES. Accordingly, it
has been the focus of many recent investigations evaluating the needed capacity of EES, of thermal storage and of additional dispatchable 
productions to help absorb large penetrations of new RES~\citep{schill2014residual,saarinen2015power,ueckerdt2015representing,schweiger2017potential}.
In his analysis of negative price regimes, \cite{nicolosi2012economics} illustrated a connection between the residual load and the merit order.
To the best of our knowledge, \cite{roon2010modeling} have been the only ones so far to report
a direct correlation between residual load and spot electricity prices. Their investigation of the German electricity market before 2010 further assumed that 
the coal and natural gas price determine the electricity price most of the time, because the German load back then required coal and gas power plants to 
produce most of the time. They therefore proposed to model electricity prices as a function of the natural gas price and of the residual load.
Given uncertainties in future gas prices (as well as those of other fossil fuels) and the level of CO$_2$ taxes, we depart from that analysis and go one
step further by modeling prices using the residual load only.  

\subsection{Our contribution}

In this manuscript we construct a model to investigate future economic conditions in the European electricity sector. To construct our pricing algorithm, we depart from \cite{roon2010modeling} in that (i) we model electricity prices only as a function of the residual load, (ii) we
take changes in production fleet and other scenarios into account in our study, (iii) we apply our study to most European countries, as they also exhibit
high degrees of correlation between day-ahead prices and residual load, and (iv) our pricing procedure may be incorporated into an aggregated European grid \citep{pagnier2017predictive}. We call the resulting model {\it Eurotranselec}. In the period 2010--2015,
after the work of \cite{roon2010modeling}, the new RES penetration in Germany has dramatically increased, with the WT production more than doubling 
and the PV production more than tripling.
Point (i) is therefore an important and necessary departure from \cite{roon2010modeling}, because with this strong increase in new RES capacities, electricity prices are less often directly determined by natural gas prices. 
In analyzing other European countries, we moreover observe that the correlation
between residual load and day-ahead prices is generally stronger in countries with higher penetration of RES.
Because  RES is expected to significantly increase in the future, it is natural to expect that this correlation will also increase.
Additionally, day-ahead markets make a significant part of the total traded electrical energy (see Table 1 below), a share which will increase with the end of long-term 
contracts in the liberalized European market.
It seems therefore reasonable to expect that the pricing model we present in this article will 
become more and more accurate as the energy transition proceeds. 
While \cite{roon2010modeling} should get the credit for uncovering an important correlation between residual load and spot market prices,
the present manuscript uses the full analytical power behind this correlation for the first time, to the best of our knowledge. Our results allow to anticipate how 
the energy transition should proceed in its next steps in order to overcome the paradox described above.

The European electricity market is expected to evolve fast with the energy transition. New  incentives and taxes may appear, 
different financial 
products related to electricity may be introduced, gas and other fuel prices may fluctuate. All this
will modify the way electricity is both produced and consumed, and will significantly 
impact power plant revenues. Our purpose in this manuscript is however to investigate the relatively near future and see how  
financial conditions in the electricity sector will evolve in the next few years. Accordingly, our investigations deliberately assume a
European electricity market where the penetration of RES increases and thermal power plants are withdrawn, all other things remaining
constant. We stress that, to extrapolate our investigations to longer time scales, our approach needs to be revisited, for instance
by considering different scenarios and pricing parameters $\Delta p_{\mathrm{da}}$ 
and $p_{\mathrm{da}0}$, or different consumption profiles. In \ref{appendix}, we comment on how consumption profiles modified by 
active demand response could be incorporated in our model.

This manuscript is organized as follows. Section \ref{rl} defines the residual load. In Section~\ref{em} we comment briefly on electricity trading. In Section \ref{corr_coef} we show the strong correlation between residual loads and day-ahead electricity prices in European countries and show that they are stronger in 
countries with more new RES.
In Section~\ref{price_res_load} we construct a synthetic electricity price in each country considered in our model. That price is based on residual load and in Section~\ref{hydro} we use it to investigate future revenues
of various types of electricity productions. We focus on conventional dam hydroelectricity, as it is one of the most flexible, more easily dispatchable electricity production
and on pumped-hydro, which is to date the dominant EES solution for which a number of projects are however currently put on hold in Europe because of
low electricity prices. We finally discuss other productions, depending on their number of operation hours per year. Conclusions and future perspectives are given
in Section~\ref{section:concl}.

\section{Residual load and must-run}\label{rl}

The residual load is defined as the difference between the total consumption and the sum of all non-flexible productions \citep{denholm2011grid,schill2014residual,saarinen2015power}. 
Non-flexible productions include new RES and run-of-river hydro. Often neglected as non-flexible productions are 
{\it must-run} productions \citep{nicolosi2010wind}, which are defined as follows. Most thermal power plants face ramping costs to turn their
production on and off, and to avoid those costs, they keep producing even when electricity prices are below their production costs. That part of their production is what is called must-run.
It is consistent with the definition and meaning of the residual load to include must-run productions in non-flexible productions and treat them as demand reduction. 
The residual load $R_i$ is then defined in each country/region (labeled by an index $i$) in our model  as
\begin{equation}\label{eq:rc1}
R_i(t)=L_i(t)-P_i^{\mathrm{PV}}(t)-P_i^{\mathrm{WT}}(t)-P_i^{\mathrm{MR}} \, .
\end{equation}
Here, $L_i(t)$ is the regional consumption/load, and $P_i^{\mathrm{PV}}(t)$, $P_i^{\mathrm{WT}}(t)$ and $P_i^{\mathrm{MR}}$ are PV, WT and must-run productions respectively. In this manuscript, they are taken at discrete times $t=n\Delta t$, with $\Delta t=1$ hour. 
In a given year, $P_i^{\mathrm{MR}}$ does not depend on time.

We take $L_i(t)$ as the 2015 consumption from \cite{entsoe2015transparency} without modification, given the relatively short time span
of our investigations in this manuscript. For investigations further into the future, other consumption profiles, and other 
consumption curves (for instance modified by active demand response, see \ref{appendix}) can be loaded into {\it Eurotranselec}. PV and WT 
productions are obtained from \cite{entsoe2015transparency}, 
which we rescale country by country to take into account planned capacity evolution as given in \cite{entsoe2015tyndp}. 

\begin{figure}[!h]
\center
\includegraphics[width=\columnwidth]{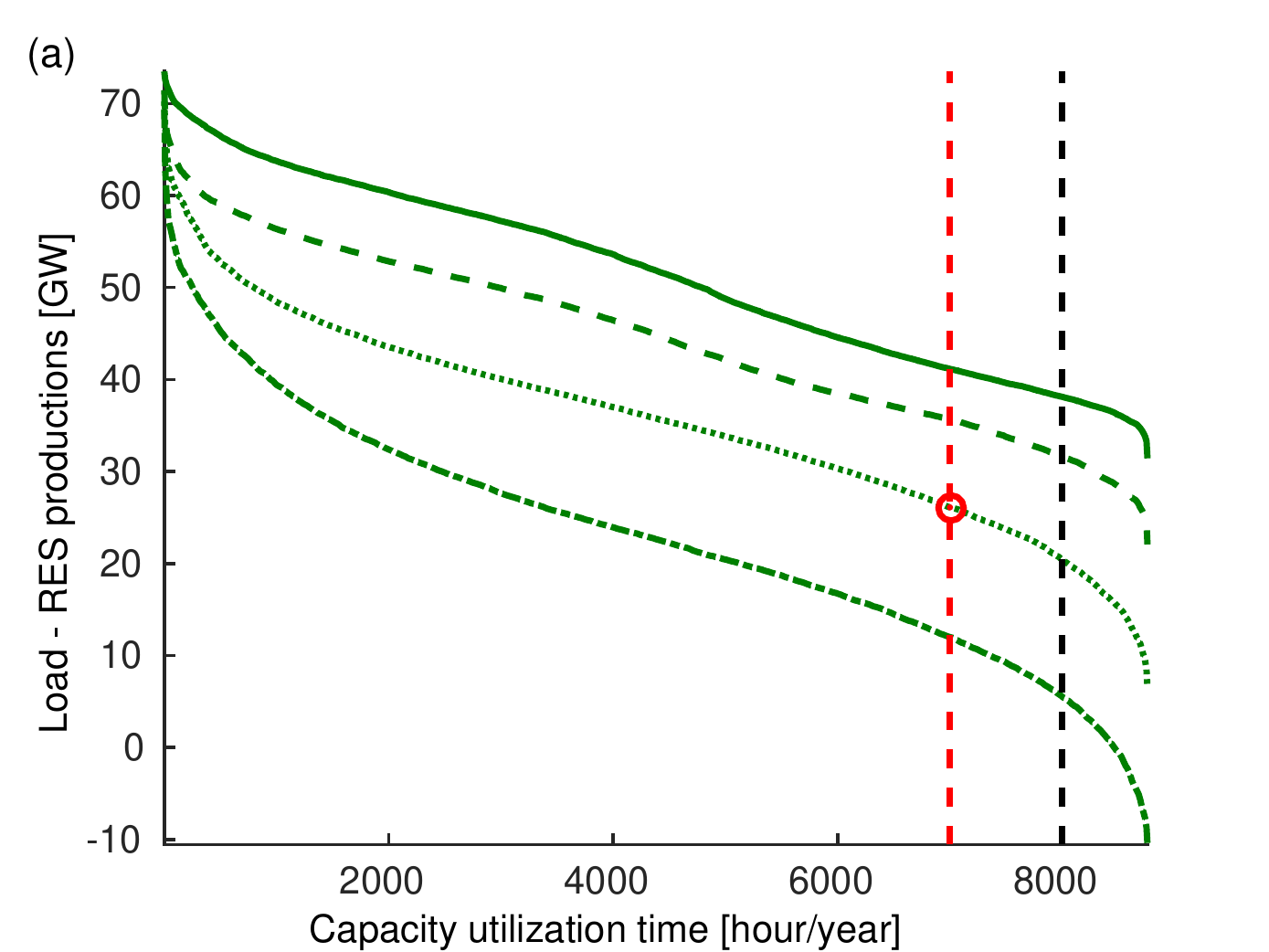}
\includegraphics[width=\columnwidth]{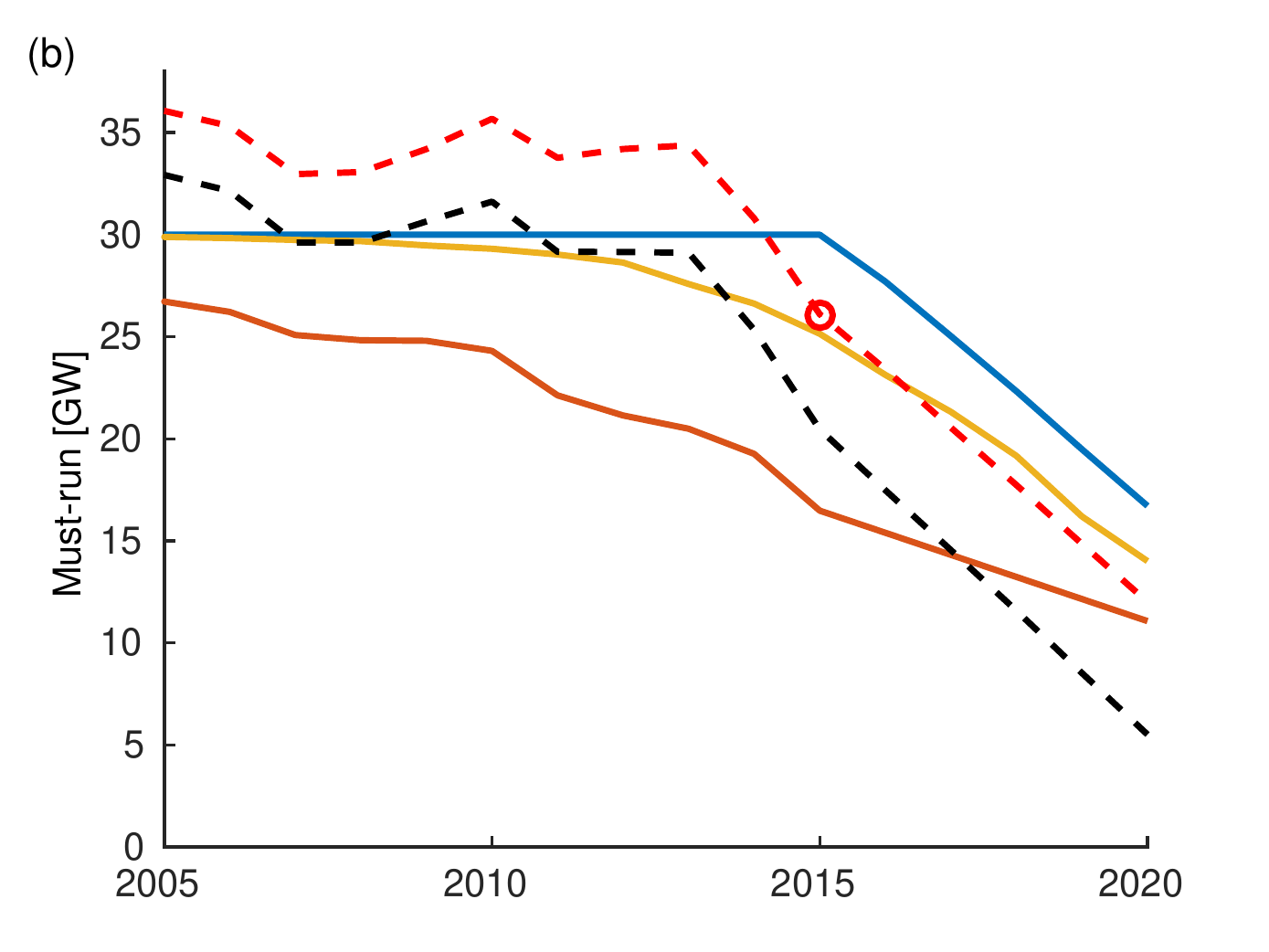}
\caption{(a) Duration curves of German load minus RES productions for the years 2000 (solid), 2010 (dashed), 2015 (dotted) and 2020 (dash-dotted). (b) Must run power 
as obtained from the duration curves [red and black dashed curves, corresponding to the red/black dashed vertical lines in panel (a)] and our three scenarios: 
keeping thermal production capacity "as long as possible" (blue), "exact substitution" of thermal production with new RES production (red), and 
smooth, in-between "interpolated path" (orange). The red circles illustrate the connection between panel (a) and panel (b).} \label{ldc}
\end{figure}

To obtain $R_i(t)$, we are left with evaluating the
must-run power which is not a uniquely defined procedure \citep{schill2014residual,denholm2011grid}. To do so, 
we evaluate the must-run from duration curves which give the number of hours in a year that a given load is exceeded.
Fig.~\ref{ldc}~(a) shows duration curves for the total consumption minus the total RES production for four different years in Germany. We extract the must-run as the 
corresponding power threshold exceeded during "most of the year", and chose this to mean 7000 (vertical red dashed line) or 8000 hours (black dashed line). 
The obtained must-run is plotted in Fig.~\ref{ldc}~(b) for these two choices (dashed lines). We see that the two curves mostly differ by a vertical shift of 3-4 GW. 
The must-run is about 30-35 GW in 2010, and keeps decreasing thereafter, as the penetration of RES increases and thermal plants are phased out. 
%%%%%%%%
%%%%%%%%

This is not the only possible procedure to evaluate the must-run but it
agrees well with another, altogether different method. \cite{nicolosi2010wind} plots electricity prices hour by hour as a function of  
the percentage of the used production capacity for various types of production in Germany, from October 2008 to November 2009. The resulting cloud is rather 
elongated in all cases -- and a linear regression 
is qualitatively representative of the data. From this linear regression, we may define the must-run as the capacity still used when this linear regression intersects
the horizontal axis between positive and negative prices. One obtains 
a must-run corresponding to $85\%$ of the total nuclear capacity, $70\%$ of the total capacity of lignite power plants and $10\%$ of the total capacity of hard coal
power plant. This estimate sums up to about 30-35 GW for 2010 in Germany, 
in agreement with our estimate extracted from duration curves. We therefore validate our procedure for estimating the volume of must-run
production and use
it to compute residual loads based on the scenario {\it 2020 Expected progress} of \cite{entsoe2015tyndp}. 

It is important to realize that the procedure just described underestimates (overestimates) the must-run for exporting (importing) countries. As a matter of fact, 
Fig.~\ref{ldc} (b) suggests that the German must-run started to decrease already in 2013, instead, 
Germany's thermal production capacity has been kept constant in 2013--2016 while its exports have increased significantly. This suggests that Germany will keep 
a large must-run as long as it can export its production when needed. To take this effect into account, we introduce three different scenarios for must-run evolution which we will
use in our investigations. These three scenarios are shown in blue, orange and red in Fig.~\ref{ldc} (b).
The blue curve corresponds to a must-run that is constant until 2015 after which it decreases with the same rate of $3$ [GW/year] in 2016-2020
as the dashed lines.
The red curve corresponds to the opposite case where thermal capacities are withdrawn exactly at the same rate 
as new RES are installed. Finally, the orange curve is a smooth curve interpolating somehow arbitrarily 
between the blue and red scenarios. Below, we dub these three scenarios {\it as long as possible} (blue), {\it interpolated path} (orange) and {\it exact substitution} (red).
We use these scenarios to investigate what ingredient(s) determine(s) the evolution of electricity prices. None of them is actually realized, however investigating the three of them allows to understand quantitatively the influence of must-run on electricity prices.

While we just focused on the German case to describe the procedure for evaluating must-run capacity, 
the described method is applied below to other European countries.

\section{Electricity day-ahead markets}\label{em}
In the liberalized European electricity market, day-ahead transactions correspond to a significant share of the total consumed electricity. This is shown 
in Table~\ref{da_table}. This share is expected to keep increasing in the future as the remaining long-term contracts expire, presumably 
with at most partial renewal because of market
liberalization. In the next section we show the strong correlation between the residual load and the day-ahead price, which allows us to model future day-ahead prices.
Given the sizeable share of day-ahead transactions, a share that will keep increasing in the coming years, we argue that this model gives us a faithful, qualitative model for
future electricity prices (not only day-ahead).

\begin{table}[!h]
\center
\begin{tabular}{lrr}
 \hline
  Market zone&Traded energy& Load percentage\\
   &\multicolumn{1}{l}{[TWh]} &\multicolumn{1}{l}{[\%]}\\
  \hline
  AT\hspace{1pt}/\hspace{1pt}DE& 264& 53\\
  BE& 24& 29\\
  CH& 23& 38\\
  CZ& 20& 28\\
  ES\hspace{1pt}/\hspace{1pt}PT& 259& 79\\  
  FR& 106& 23\\
  IT& 195&  62\\
  NL& 43& 39\\
  NO& 133& 103\\
  PL& 24& 18\\
  SE& 128& 94\\
  UK& 47& 19\\
  \hline 
  Total & 1264 & 49 \\
  \hline
\end{tabular}
\caption{Traded electrical energy and corresponding load percentage of several European day-ahead markets in 2015. Sources: \cite{epex2015annual}, \cite{omie2015main}, \cite{ote2015annual} and \cite{nordpool2015annual}.}\label{da_table}
\end{table}

\section{Correlations between residual loads and day-ahead prices}\label{corr_coef}

Fig.~\ref{def_rl} illustrates the strong correlation between national residual loads $R_i(t)$ and day-ahead prices $p_{\mathrm{da}i}(t)$, a correlation that had been noticed 
by \cite{roon2010modeling} for Germany in 2007--2009. 
In this section, we further quantify this correlation for other European countries. 
Statistical correlation between discrete sets of data $X=\{x_k\}$ and $Y=\{y_k\}$ is standardly measured by Pearson's correlation coefficient
\begin{equation}\label{eq:coef}
r(X,Y)= \frac{\sum_{k=1}^{n}(x_k-\bar x)(y_k-\bar y)}{\sqrt{\sum_{k=1}^{n}(x_k-\bar x)^2\sum_{k=1}^{n}(y_k-\bar y)^2}},
\end{equation}
where $\bar x$ and $\bar y$ are the average values of the two sets. By definition, one has
 $r \in [-1,1]$, with $r=0$ indicating the absence of correlation between the two sets, $r=1$ two perfectly correlated sets and $r=-1$ two totally anticorrelated sets. A value 
$r>0.5$ indicates an already strong correlation. 

We calculate Pearson's coefficient for different years and different European countries, based on their residual loads constructed as described in Section~\ref{rl}. For each reported year, we used hourly sets of data for both residual loads and day-ahead prices, from
which we removed the 2 \% highest and lowest values -- corresponding to those that are further away from the average than between two and three standard deviations.  
These extreme events correspond to exceptional situations
with forecast errors, unplanned production outages and so forth \citep{creg2015price,christensen2012forecasting}, events that are hardly predictable and lay
beyond the scope of the present manuscript. 
Table~\ref{corr_hist} shows the evolution of correlations between residual loads and day-ahead prices of the four largest countries in continental Europe.
All values are large, $r>0.58$, and seem to be constant or perhaps even increasing with time

\begin{table}[!h]
\center
\begin{tabular}{lrrrr}
\hline
$r(p_{\mathrm{da}i},R_i)$& 2012& 2013& 2014& 2015\\
\hline
FR& 0.65 &0.74 &0.71 &0.67\\	
DE& 0.78 &0.86 &0.89 &0.89\\
IT &0.63 &0.58 &0.61 &0.77\\
ES &\# &\# &0.77 &0.88\\
\hline
\end{tabular}
\caption{Evolution of the correlation between national day-ahead prices and residual loads.}\label{corr_hist}
\end{table}

\begin{figure}[!h]
\center
\includegraphics[width=\columnwidth]{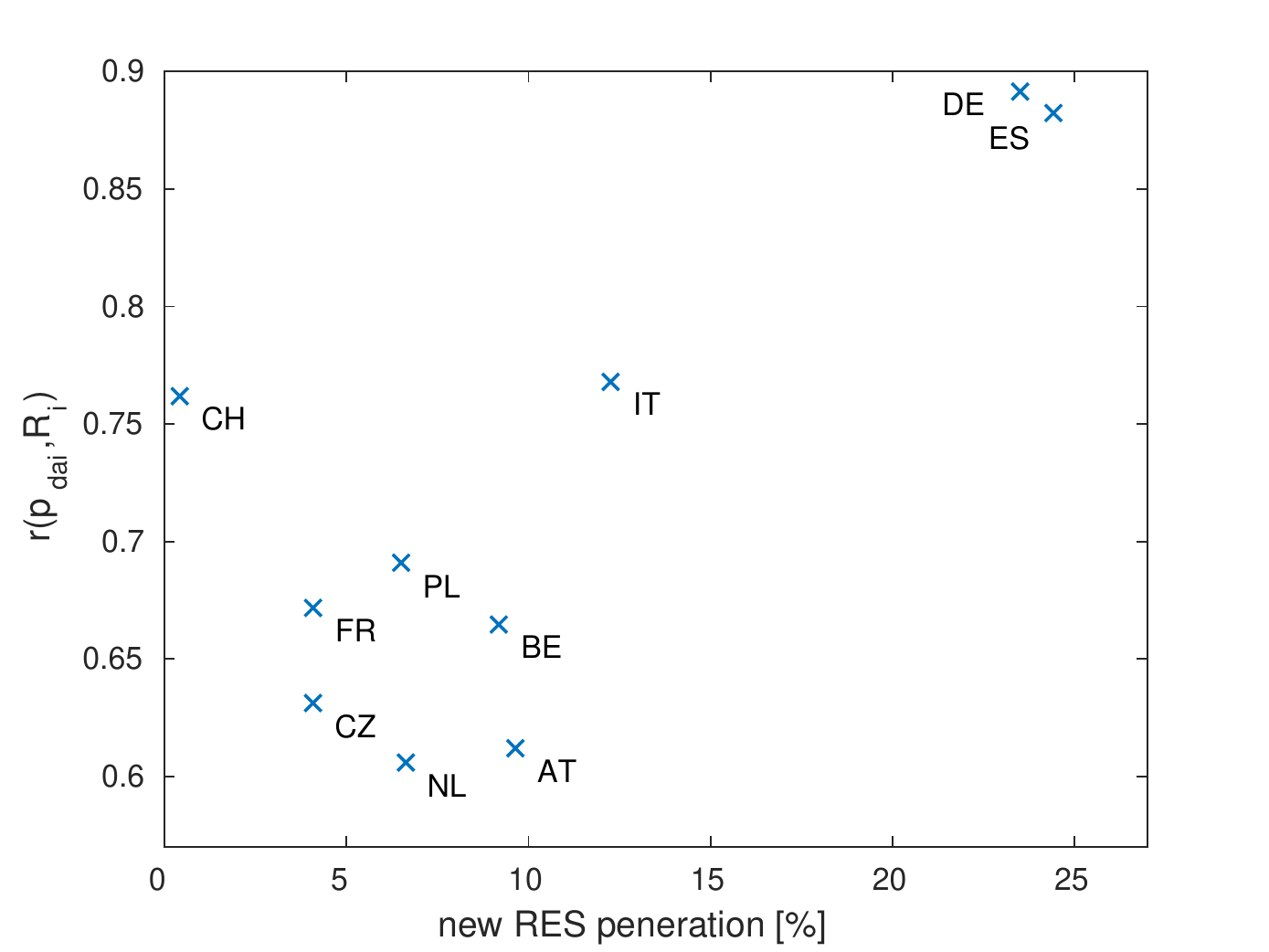}
\caption{Pearson's correlation coefficient $r$ between national residual loads $R_i$ and 2015 national day-ahead prices $p_{\mathrm{da}i}$ as a function of 
the penetration of new RES in several European countries.
}\label{corr}
\end{figure}

We further investigate the correlation coefficient for 2015 data in a number of continental European countries.
Fig.~\ref{corr} plots the correlation coefficient between national residual load and day-ahead electricity price as a function of new RES penetration, which we
took as the ratio of yearly RES production to the total electricity production. Data are taken from \cite{entsoe2015transparency} and have been crosschecked and completed where necessary  with data obtained from national grid operators and power markets.\footnote{The Italian electricity market has regional prices and we chose to use the Northern 
Italy price as national price in Fig.~\ref{corr}, giving $r=0.768$. We found similar values 
of $r=0.7641$ and $0.7506$ for the Central Northern and Central Southern prices respectively and a slightly weaker correlation with $r=0.6869$ for the 
Southern Italy price.}

The correlation coefficients in Table~\ref{corr_hist} and Fig.~\ref{corr} satisfy $r(p_{\mathrm{da}i},R_i) >0.58$ in all cases, indicating a strong correlation between the residual loads and day-ahead prices. 
Additionally, $r(p_{\mathrm{da}i},R_i)$ is larger in countries with larger penetration of new RES (see Fig.~\ref{corr}), with the exception of Switzerland, where the correlation is 
presumably higher due to a large penetration of hydroelectricity - an "old" RES. Given this trend, and the planned increase in new RES penetration in all European countries, it seems
natural to expect an even larger correlation between residual loads and day-ahead prices in the future.

\section{Present and future electricity prices modeled after the residual load}\label{price_res_load}

We argued in Section~\ref{corr_coef} that the already large correlation factor $r$ between day-ahead prices and residual loads (see
 Table~\ref{corr_hist}) is expected to become even larger
as  the penetration of new RES increases. This is so, because, as is illustrated in Fig.~\ref{corr}, $r$ is 
larger in countries with larger amounts of new RES. Simultaneously, day-ahead markets represent a significant part of all electricity
transactions as is shown in Table~\ref{da_table}, a share that is likely to keep increasing in Europe as the liberalization of the market becomes complete. It therefore makes sense to model future electricity prices from residual loads. 
The latter are based solely on scenarios for the evolution of the consumption, the future RES productions
and the must-run. The economic feasibility and the future of different production types under given scenarios can then be checked quantitatively. 
In this section we construct such a price 
and show that it reproduces historical prices with very good accuracy, except for rare extreme events. 

\begin{figure}[!h]
\center
\includegraphics[width=\columnwidth]{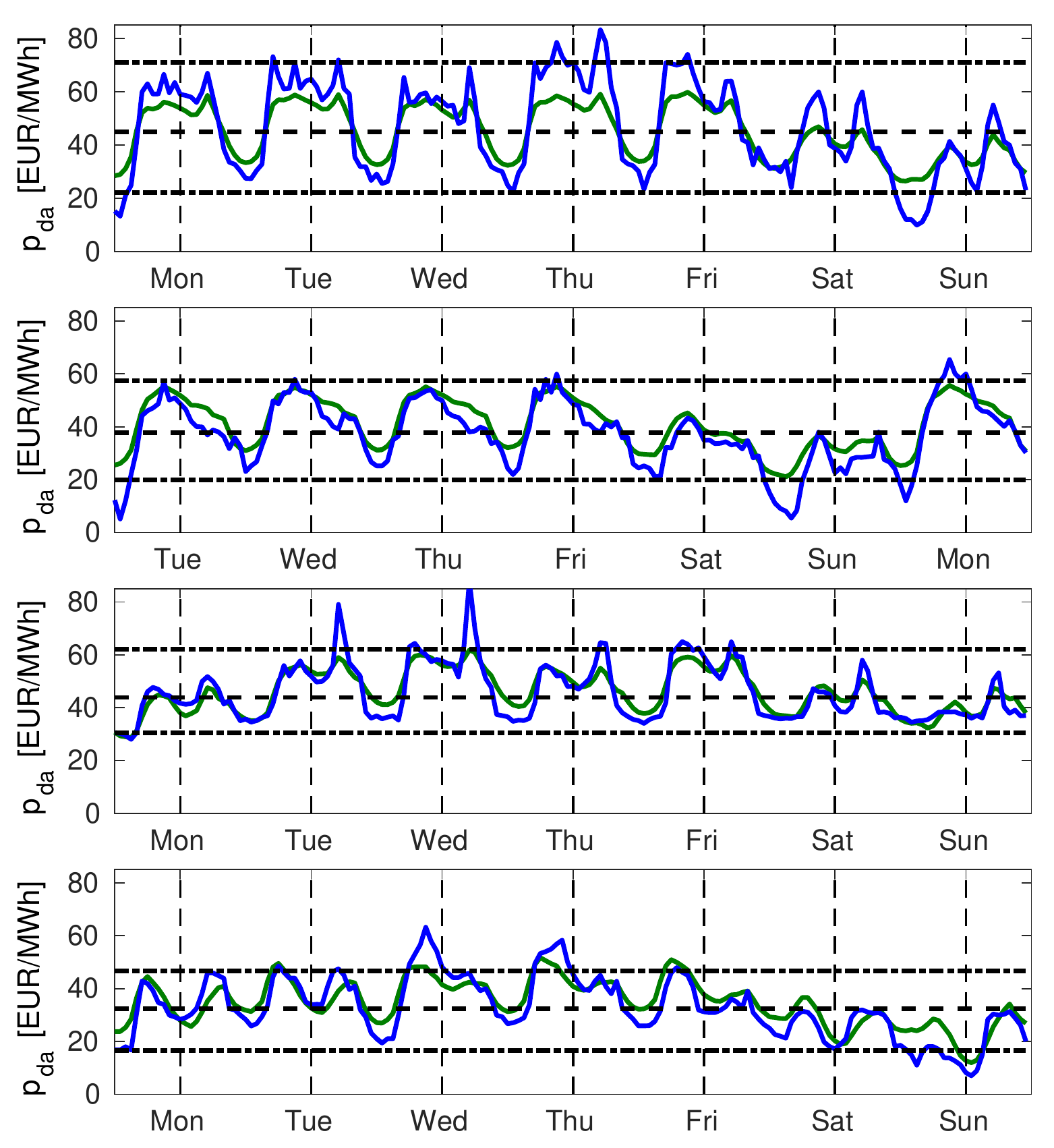}
\caption{Electricity price \eqref{pda} built on the residual load (green line) and actual day-ahead electricity price (blue line) 
during a week in winter and summer 2006 (top two panels) and 2013 (bottom two 
panels) in Germany. Dashed lines show the monthly average price and the dotted-dashed lines prices exceeded 10 and 90 \% of the time
during that month. Vertical dashed lines indicate noon time.}\label{price}
\end{figure}

We construct a synthetic electricity price as a linear regression of the residual load,
\begin{equation}
p_{\mathrm{da}}(t)=\Delta p_{\mathrm{da}}R(t)+p_{\mathrm{da}0} \, ,\label{pda}
\end{equation}
where we rewrote Eq.~\eqref{eq:lin_correl}. For the sake of simplicity, we drop the country index $i$ here. We focus on electricity prices and revenues for various
productions in Spain and Germany, two large European countries that are already well engaged in their energy transition in the electric sector, with large penetration 
of new RES. The RES mix has proportionally less PV in Spain than in Germany, which allows us to identify  differences in the evolution of prices from different
choices of RES mixes. Based on 2015 data, we obtain $\Delta p_{\mathrm{da}}\approx 1$ and $2.2$ [\euro{}/MWh$\cdot$GW$^{-1}$] and $p_{\mathrm{da}0}\approx 20$ and $30$ [\euro{}/MWh] in Germany and Spain respectively for the parameters in Eq.~\eqref{pda}. We found very little
change in these parameters during the years 2013--2015 in Germany and therefore assume these parameters to be constant in time 
in each country, for a time window ranging from 2015 to 2020. 
That this is reasonable is illustrated in Fig.~\ref{price} which shows that the effective price $ p_{\mathrm{da}}(t)$ of Eq.~\eqref{pda}
reproduces historical day-ahead prices quite well. The agreement is already good in 2006 and becomes even better in 2013. 
Exceptional price spikes and troughs are not totally captured, which correspond however to unusual situations
beyond the reach of our modeling.

\begin{figure}[!h]
\center
\includegraphics[width=\columnwidth]{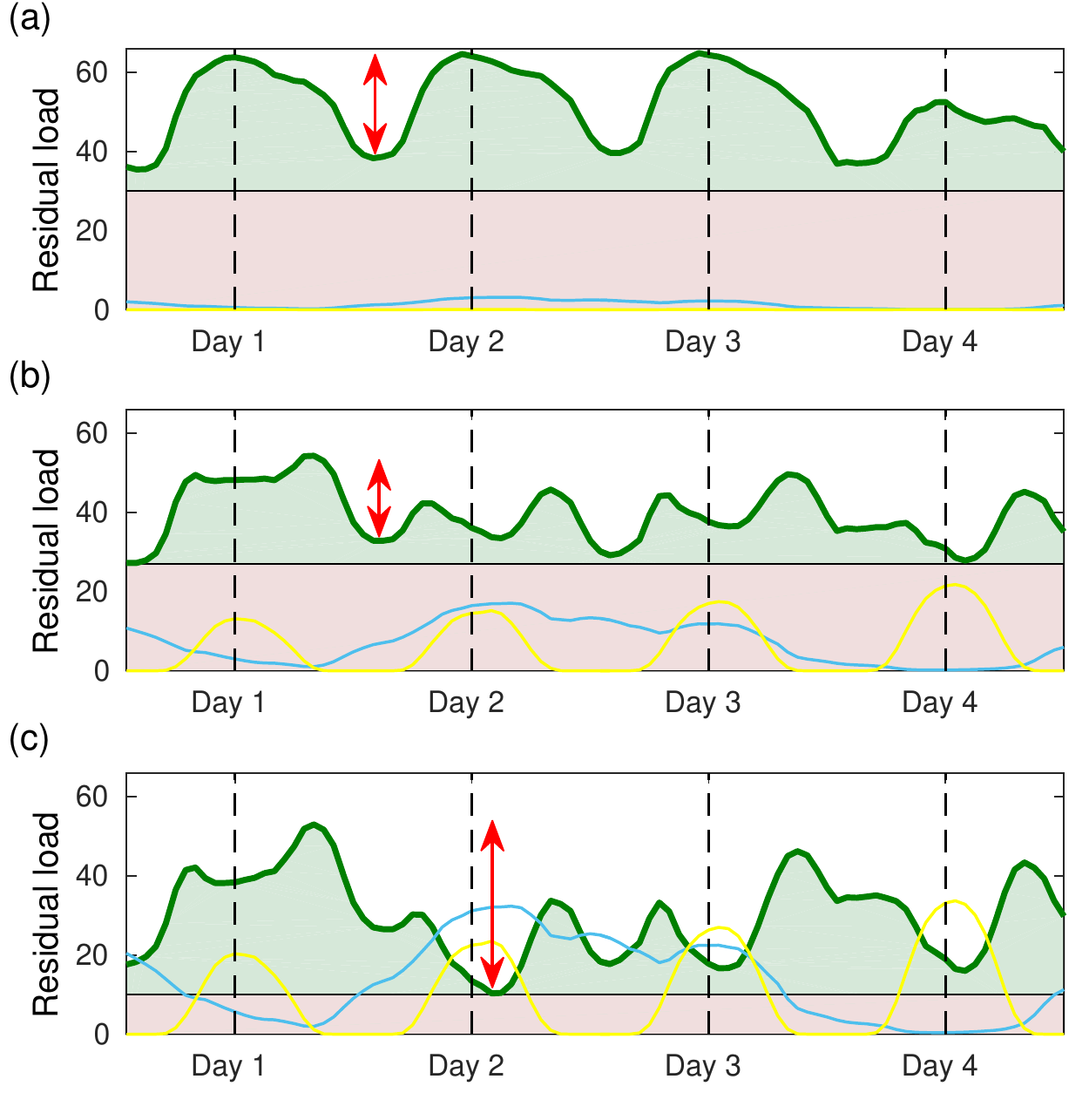}
\caption{Sketch of the residual load (green area) and must-run (light red band) at three stages of the energy transition: (a) Initial, (b) intermediate and (c) 
late stages of the transition. PV (yellow) and WT (light blue) production profiles are superimposed. Red arrows indicate the magnitude of fluctuations of the residual load.
Vertical dashed lines indicate noon time.}\label{sketch}
\end{figure}

Eq.~\eqref{pda} allows us to qualitatively forecast  
electricity prices and price fluctuations in the framework of the energy transition. 
The latter substitutes thermal productions with new RES. Doing so, it reduces the must-run and changes fluctuations in the residual load. 
The way these fluctuations change depends on the chosen RES mix: PV produces more around noon, it therefore is correlated with the main load peak;
WT production on the other hand is effectively random in time on time scales of the order of few hours to few days, and therefore 
uncorrelated with consumption on such time scales.
Consequently, fluctuations in residual loads will always increase if the substitution mix is made of 
WT only, while they will first decrease before increasing again if the mix is dominated by PV. This is illustrated in Fig.~\ref{sketch} which sketches the
behavior of the
residual load at three different stages of the energy transition. Panel (a) shows the situation at the very initial stage 
of the energy transition, with low RES penetration. The shape of the residual load is very similar to the load itself and the must-run is high. 
Panel (b) illustrates the transition period with increased RES penetration with a significant fraction of PV, corresponding to the German mix. PV significantly decreases 
the load peak during office hours, which reduces fluctuations of the residual load. The must-run is still high. In our model, this reduces fluctuations in electricity prices, 
therefore there are less financial opportunities for flexible productions.
In the final stages of the energy transition, the large RES penetration completely changes the shape of the residual load, which looks now very different from 
the load, see Fig.~\ref{sketch}~(c). The must-run power is lower, bringing average prices higher.
Most importantly, fluctuations in the residual load are comparable to and even higher than those at the early stages of the energy transition.

The residual load quantifies the balance between non-flexible supply and demand, and accordingly, 
the correlation between electricity prices and residual load can be understood as a logical consequence of the economic law
of supply and demand. This correlation may vary in the future, however, from fundamental laws of economics, it is likely to
remain sizeable in any event. It therefore makes sense to introduce an electricity price as in Eq.~(\ref{pda}). Varying economic
conditions may however impact the pricing parameters $\Delta p_{\mathrm{da}}$ and $p_{\mathrm{da}0}$ and there is no reason 
a priori to consider them constant in time (the scenario we discuss in this manuscript). Other scenarios 
with varying $\Delta p_{\mathrm{da}}$ and $p_{\mathrm{da}0}$ can be investigated. Qualitatively, one anticipates that  $p_{\mathrm{da}0}$
is determined by the marginal cost of must-run production. In Europe this is essentially the marginal price of electricity from coal-fired plants, 
and therefore $p_{\mathrm{da}0}$ increases if carbon taxes increase. The parameter $\Delta p_{\mathrm{da}}$ on the other hand is more 
directly related to the order of merit, and thus to the marginal cost of electriciy from gas-fired power plants. As such it will follow the evolution
of both carbon taxes and natural gas prices. How much these parameters vary for given variations in gas prices and carbon taxes needs to 
be calibrated.
Performing this calibration goes beyond the purpose of the present manuscript and is left to future works. Here, we consider
constant $\Delta p_{\mathrm{da}}$ and $p_{\mathrm{da}0}$ and restrict our investigations to a relatively short time
window, until 2020.

With these considerations, and under the assumptions described above, it is easy to qualitatively predict the evolution of electric revenues.
Consider for instance a high-power pumped-storage (PS) hydroelectric plant. 
Its revenues directly depend on the difference between highest and lowest prices. From the 
discussion above, a PS plant sees its revenue first decrease in the initial stages of the energy transition, where increased RES penetration reduces
price fluctuations. The revenue however increases later, once the RES penetration is such that it restores large fluctuations in residual loads and thus in electricity prices. 
In the upcoming sections we show that the intermediate period of reduced revenues for flexible productions depends on (i) the rate
at which RES penetration increases, (ii) the chosen RES mix, and (iii) the rate at which must-run is reduced. To overcome the paradox of the energy
transition, one needs to chose scenarios such that these three ingredients, when combined, reduce the duration of the intermediate period with low revenues.

\section{Future revenues by electricity production type}\label{hydro}

We investigate the future revenues of different electricity productions in Europe with the synthetic electricity price of Eq.~\eqref{pda}. We initially focus on the hydroelectric sector, which 
can provide flexibility of production and storage capacities needed to integrate new RES into the electric grid. We next turn our attention to general power plants characterized by their annual number of operation hours.

\begin{figure}[!h]
\center
\includegraphics[width=\columnwidth]{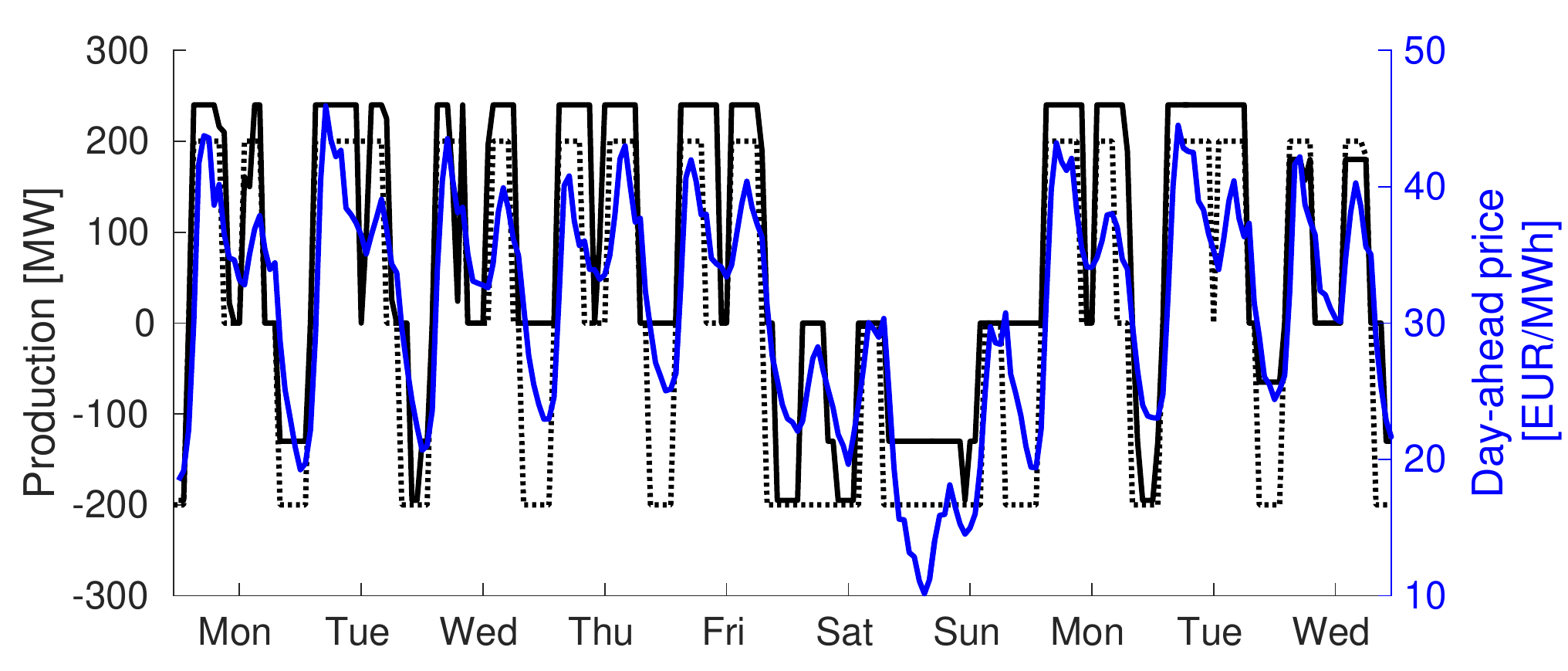}
\caption{The actual (solid black curve) and computed (dotted black curve) production of a typical Swiss PS plant and the day-ahead price (blue curve). 
Negative production means pump load. Sources: Swissgrid and \cite{entsoe2015transparency}.}\label{PS_op}
\end{figure}

\subsection{Revenues of a pumped-storage plant and the fluctuations of the residual load}
The revenue $G$  of a PS plant over a time interval $t \in [t_i,t_f]$ is given by
\begin{equation}
G = \int_{t_i}^{t_f} p_{\mathrm{da}}(t)\cdot P_{\mathrm{PS}}(t)\mathrm{d}t,\label{ips}
\end{equation}
where $P_{\mathrm{PS}}(t)$ is the electric power produced ($P_{\mathrm{PS}}(t)>0$) or consumed ($P_{\mathrm{PS}}(t)<0$) by the plant. 
Optimizing the revenue of PS plants means producing when $p_{\mathrm{da}}(t)$ is large and consuming when $p_{\mathrm{da}}(t)$ is low. It therefore
makes sense to assume that $P_{\mathrm{PS}}$ and $p_{\mathrm{da}}$ are strongly correlated. This assumption is confirmed in Fig. \ref{PS_op}, which  
shows the production of a Swiss PS plant and the day-ahead price for 10 consecutive days in 2015. Neglecting losses for the time being, we write 
\begin{equation}
P_{\mathrm{PS}}(t)\cong \pi_{\mathrm{PS}}\cdot [p_{\mathrm{da}}(t)-\bar p_{\mathrm{da}}],\label{eq_ps}
\end{equation}
where $\bar p_{\mathrm{da}}$ is the average price over the considered time period 
and $\pi_{\mathrm{PS}}$ is a prefactor linking prices to production. Eq.~\eqref{eq_ps} guarantees that $\int_{t_i}^{t_f}P_{\mathrm{PS}}(t)\mathrm{d}t=0$ as should be for a PS plant without loss.
Inserting Eq.~\eqref{eq_ps} into \eqref{ips} and using Eq.~\eqref{pda}, we get 
\begin{equation}\label{eq:revenues}
G \cong \pi_{\mathrm{PS}} \, \Delta p^2_{\mathrm{da}} \, T \, \mathrm{Var}[R] \, ,
\end{equation}
for the annual revenue with $T=8760$ hours.
This result shows that the revenue of a lossless PS plant is proportional to the variance of the residual load. Eq.~\eqref{eq:revenues} formalizes the relation qualitatively discussed at the end of Section 5 between revenues of PS plants and fluctuations of the residual load.

\begin{figure}[!t]
\center
\includegraphics[width=\columnwidth]{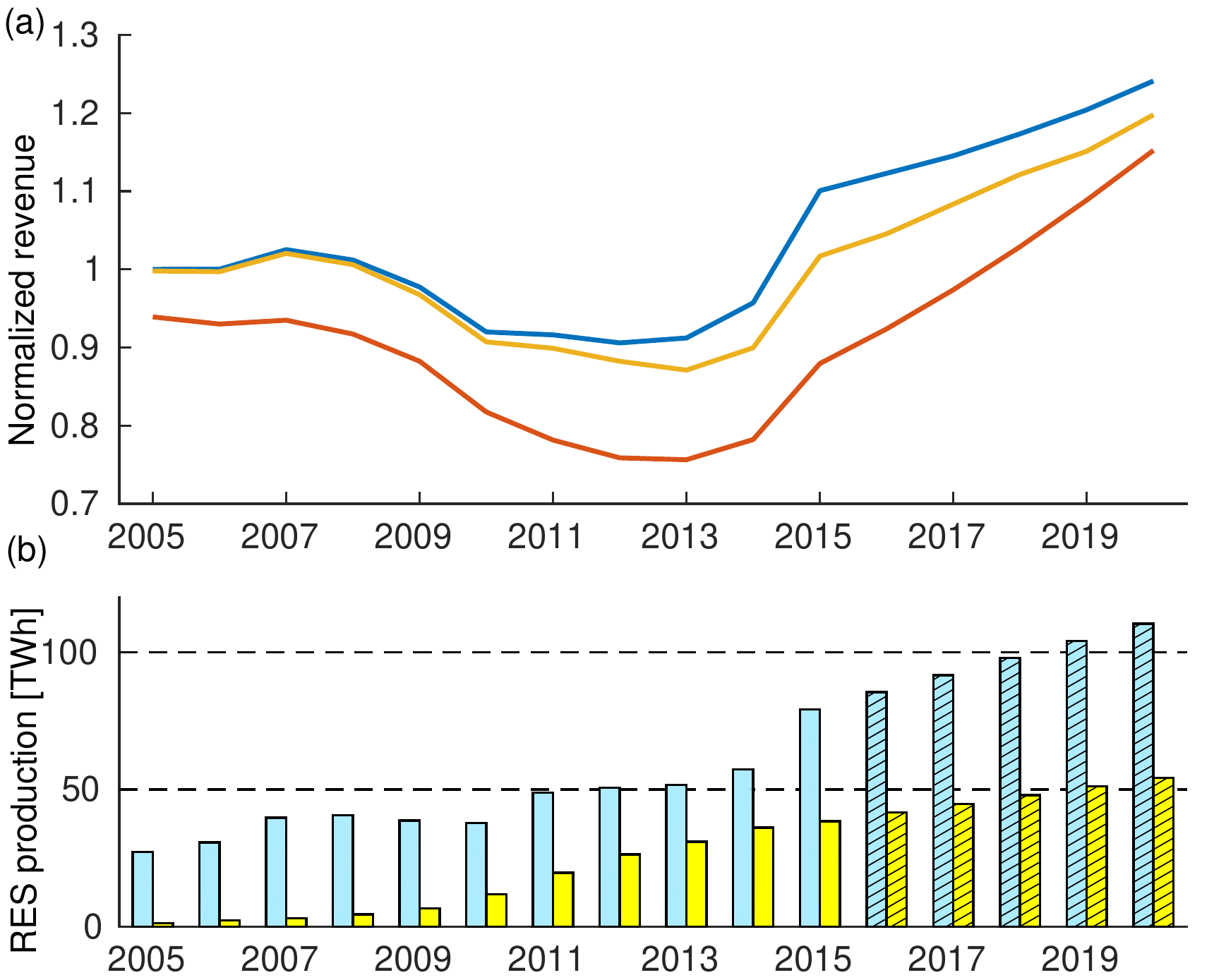}
\caption{(a) Normalized revenue (divided by the revenue of 2005 in the scenario "as long as possible") of a German PS plant with $\eta=0.9$,
 for different scenarios of must-run withdrawal (see Section~\ref{rl}) : "as long as possible" (blue), "interpolated path" (orange) and "exact substitution" (red). 
 (b)  Evolution of WT (light blue) and PV (yellow) annual production in Germany. Dashed rectangles correspond to planned future evolution~\citep{entsoe2015tyndp}.}\label{revenue_PS}
\end{figure}

\subsection{Future revenues of a pumped-storage plant}

\begin{figure}[!t]
\center
\includegraphics[width=\columnwidth]{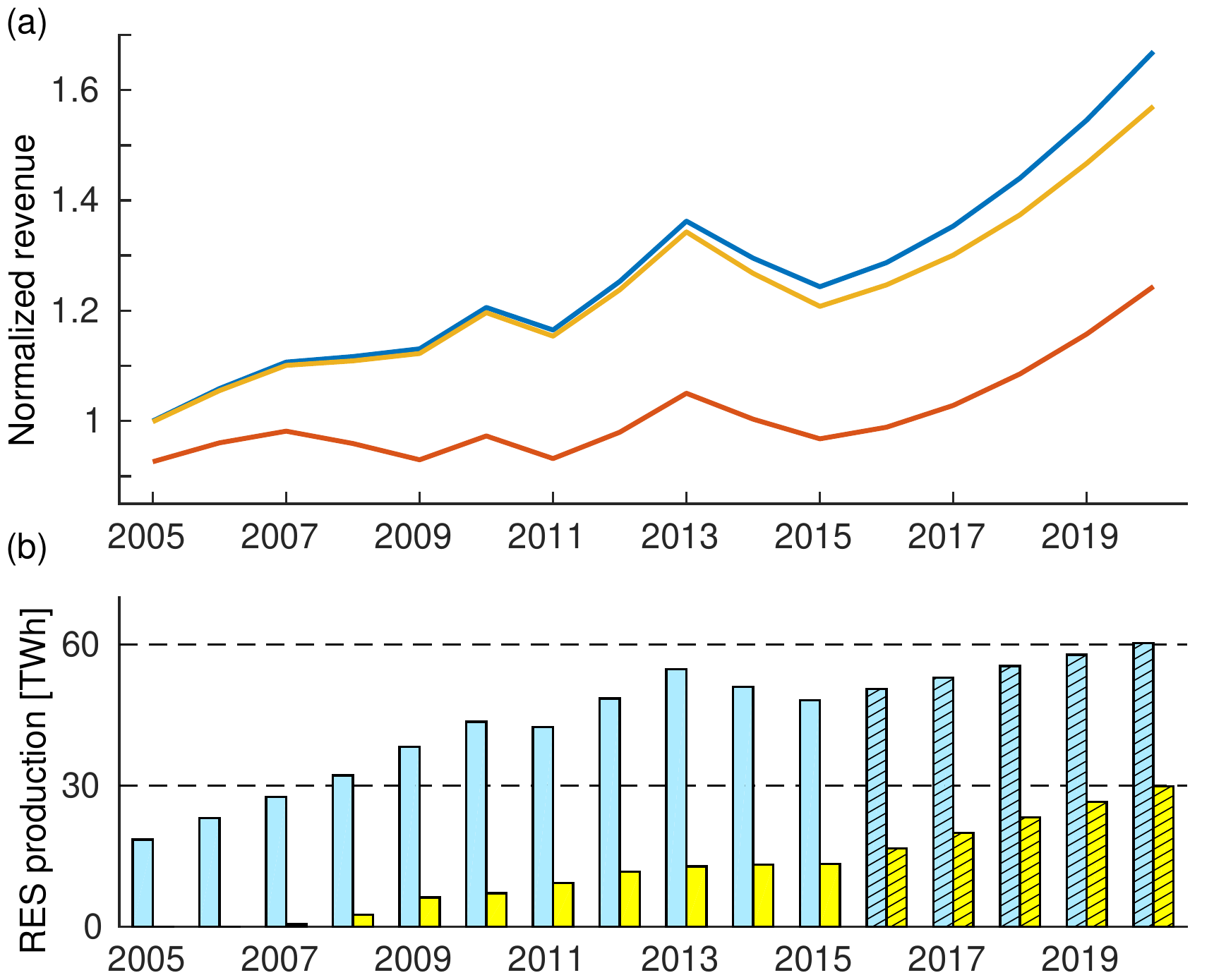}
\caption{(a) Normalized revenue (divided by the revenue of 2005 in the scenario "as long as possible") of a Spanish PS plant with $\eta=0.9$,
 for different scenarios of must-run withdrawal (see Section~\ref{rl}) : "as long as possible" (blue), "interpolated path" (orange) and "exact substitution" (red). 
 (b)  Evolution of WT (light blue) and PV (yellow) annual production in Spain. Dashed rectangles correspond to planned future evolution~\citep{entsoe2015tyndp}.}\label{revenue_ES} 
\end{figure}

We next investigate numerically the revenue of PS plants in Germany and Spain in the period 2005-2020. Residual loads are calculated from 
2015 data for the load and RES profiles, the latter being scaled up from year to year to interpolate linearly between the 2015 realized annual production and the planned
2020 annual production~\citep{entsoe2015tyndp}. The evolution of the annual RES productions is given in Figs.~\ref{revenue_PS}~(b) and \ref{revenue_ES}~(b). The revenue is given by
\begin{align}
G& =\sum_k p_{\mathrm{da},k} \, P_{\mathrm{PS},k} \, \Delta t
= \sum_k p_{\mathrm{da},k}[P_{\mathrm{t},k}-P_{\mathrm{p},k}] \, \Delta t \, ,\label{eq_gain} 
\end{align}
with the time step $\Delta t$ used in the calculation -- one hour in our case -- and
where we introduced $P_{\mathrm{PS},k} = -P_{\mathrm{p},k}$ when $P_{\mathrm{PS},k}<0$ and $P_{\mathrm{PS},k} = 
P_{\mathrm{t},k}$ when $P_{\mathrm{PS},k}>0$.
The power profile is related to the evolution of the reservoir level $S_{\mathrm{PS},k}$, and 
for a PS plant with pump/turbine efficiency $0\le\eta\le 1$ each way, this relation reads
\begin{equation}
S_{\mathrm{PS},k+1} = S_{\mathrm{PS},k} + [\eta P_{\mathrm{p},k}-\eta^{-1} P_{\mathrm{t},k}]\Delta t \, .\label{eq_storage2}
\end{equation}
Finally, the reservoir level must be positive but smaller than its maximal level at all times, giving the constraint 
\begin{align}
 0\le S_{\mathrm{PS},k}\le S_{\mathrm{PS}}^{\max}\, ,  \hspace{10pt} \forall k .\label{eq_storage1}
\end{align} 
Eqs.~\eqref{eq_gain}--\eqref{eq_storage1} govern the production of a PS plant. We generate PS power profiles by maximizing the revenue $G$ of Eq.~\eqref{eq_gain}
under the constraints of Eqs.~\eqref{eq_storage2} and \eqref{eq_storage1}. Using the actual day-ahead price, this procedure generates a fictitious PS production 
profile given by the dotted black curve in 
Fig.~\ref{PS_op}, which is very close to the actual one (solid black curve). We attribute the few discrepancies to the fact that our maximization of $G$ in Eq.~\eqref{eq_gain}
is made with perfect advance knowledge of the load and RES productions. This test substantiates our procedure for calculating power profiles 
and evaluating revenues of PS plants.

We calculate revenues of PS plants from Eqs.~\eqref{eq_gain}--\eqref{eq_storage1} with the synthetic price of Eq.~\eqref{eq:lin_correl} and the usually reported efficiency of $\eta=0.9$ each way \citep{rehman2015pumped,guittet2016study}.
Revenues of German and Spanish PS plants as the energy transition unfolds are shown in Figs.~\ref{revenue_PS} and \ref{revenue_ES}. It is seen that they 
remain approximately constant in both countries from 2005 to 2008/2009, even though the WT production
increases by 40 \% in Germany and almost 50 \% in Spain. Revenues decrease significantly in Germany from 2009 on, reaching a minimum around 2013 with revenues
reduced by as much as 20 \%. The revenues decrease more for faster must-run reduction [see Fig.~\ref{revenue_PS}~(a)]. 
This is easily understandable when one realizes that a higher must-run
reduces the residual load, and with it, the electricity price. Power losses due to the finite efficiency of the plant, $\eta < 1$, cost less at higher must-run, which increases revenues.

The striking feature in Fig.~\ref{revenue_PS}~(a) is that,
as expected from the discussion in Section~\ref{price_res_load} together with Eq.~\eqref{eq:revenues}, the drop in revenues corresponds to the acceleration of the
penetration of PV, which reduces the mid-day residual load peak. The fluctuations of the residual load go down, leading to reduced revenues through Eq.~\eqref{eq:revenues}.
As the penetration of PV further increases, so do the fluctuations of the residual load -- one enters the stage depicted in Fig.~\ref{sketch}~(c) and the revenues
increase again. The drop in revenues does not last long. The importance of PV in this phenomenon becomes clear when comparing 
Fig.~\ref{revenue_PS}~(a) and Fig.~\ref{revenue_ES}~(a). The latter figure displays no
significant drop for a Spanish PS plant. 
This is so, because the mix of new RES 
is clearly dominated by WT in Spain. Fluctuations in residual load are increased at all stages of the transition,
regardless of the chosen scenario for must-run reduction. Thus, from Eq.~\eqref{eq:revenues}, revenues also always tend to increase.

\subsection{Future revenues of conventional hydroelectric plants}

We next consider conventional dam hydroelectric power plants. The main difference with PS plants is that 
(i) conventional dam hydroelectric plants only produce and (ii) their reservoir is filled by natural water inflow.\footnote{There are of course also water inflows for PS plants, however they are negligible against normal operation which typically fills and empties the reservoir in a matter of few days.} 
We model them slightly differently from PS plants.
Their revenue is given by
\begin{align}\label{eq:gaindam}
G& =\sum_k p_{\mathrm{da},k} \, P_{\mathrm{D},k} \, . 
\end{align}
We use the same 
synthetic price for $p_{\mathrm{da},k}$ in Eq.~\eqref{eq:gaindam} as for the analysis of PS plants. The power profile $P_{\mathrm{D},k}$ is related to the evolution of the reservoir level $S_{\mathrm{D},k}$, 
\begin{equation}\label{eq:leveldam}
S_{\mathrm{D},k+1} = S_{\mathrm{D},k} + [I_k-P_{\mathrm{D},k}] \, \Delta t \, , 
\end{equation}
where $I_k$ is the power corresponding to water inflow into the dam  (rain- and snowfall, snow- and icemelt) at the time interval $k$.
The reservoir level must be positive but smaller than the maximal storage capacity at all times, giving a condition similar to \eqref{eq_storage1},
\begin{align}\label{eq:constraintdam}
0\le S_{\mathrm{D},k} \le S_{\mathrm{D}}^{\max} \, , \hspace{5pt} \forall k \, .
\end{align}
As for PS power plants, we determine the power profile $P_{\mathrm{D},k}$ by maximizing the gain $G$ in Eq.~\eqref{eq:gaindam} for a typical dam hydro power plant in the
Alps. We take  $\{I_k\}$ as the water inflow averaged over all Swiss dams, as extracted from weekly dam energy content and production \citep{ofen2016weekstat}.  A conventional 
dam hydroelectric plant is characterized by its rated power $P_\mathrm{D}^{\max}$, its storage capacity $S_{\mathrm{D}}^{\max}$ and the annual energy 
inflow $E_\mathrm{D}=\sum_k I_k\Delta t$. Relative revenue evolution therefore depends on only two dimensionless parameters which we take as
$S_{\mathrm{D}}^{\max}/P_\mathrm{D}^{\max} \Delta t\equiv N_{\rm empty}$ and $E_\mathrm{D}/P_\mathrm{D}^{\max}\Delta t \equiv N_{\rm op}$, giving the number of hours of operation at full power to empty 
the reservoir and to use all the annual energy inflow respectively. 
We found that revenues depend only very weakly on $N_{\rm empty}$, and therefore focus on the evolution of revenues vs. 
$N_{\rm op}$. In multiannual average, dams annually produce their energy inflow $E_\mathrm{D}$, and in continental Europe, this usually
corresponds to $N_{\rm op} \in [1000,3000]$ hours of full power operation \citep{thuerler2014statistik}.

\begin{figure}[!h]
\center
\includegraphics[width=\columnwidth]{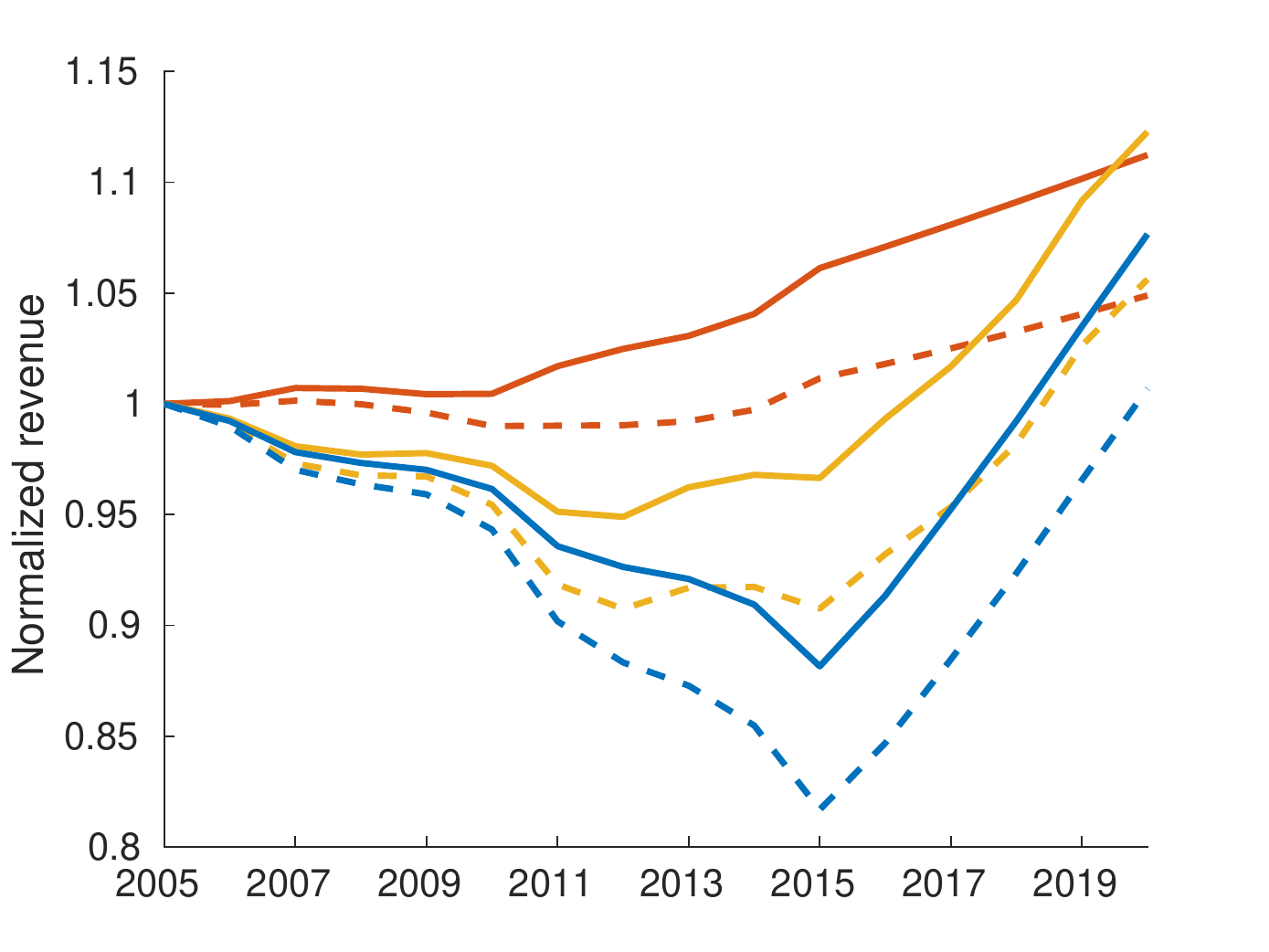}
\caption{Normalized revenue of conventional dam hydroelectric plants with high ($N_{\rm op}=1000$ hours; solid lines) and low ($N_{\rm op}=3000$ hours; dashed lines) power capacity in the Alps, with $N_{\rm empty}=1000$ hours and
for the scenarios "as long as possible" (blue), "expected path" (orange) and "exact substitution" (red) of must-run withdrawal.}\label{revenue_hydro}
\end{figure}

Fig.~\ref{revenue_hydro} shows the revenues of conventional dam hydroelectric plants with $N_{\rm empty}= 1000$ and
$N_{\rm op}=1000, 3000$. One sees a similar nonmonotonous behavior as for PS plants in Germany (Fig.~\ref{revenue_PS}).
How much and until when the revenue drops depend sharply on the 
 chosen scenario for must-run withdrawal. Generally, we find that revenues drop more and longer for delayed must-run withdrawal -- the flexibility 
 of conventional dam hydroelectric power plants is rewarded best for exact substitution of production capacity, where production from thermal plants is reduced
 at the same rate as new RES production increases. Only in that case would we see no drop in revenue for conventional hydro power plants. Regardless of the must-run scenario, 
 revenues go back to their pre-energy transition level by 2020 at the latest in all considered scenarios. We observe that plants with higher rated power $P_\mathrm{D}^{\max}$,
 i.e. lower number $N_{\rm op}$ of annual operation hours,  see their revenue decrease less than those with lower power, because the higher the rated power, the easier it is
 to produce almost only during peaks of financial opportunities.
 
 \subsection{Future revenues of power plants by annual operation time}

\begin{figure}[!h]
\center
\includegraphics[width=\columnwidth]{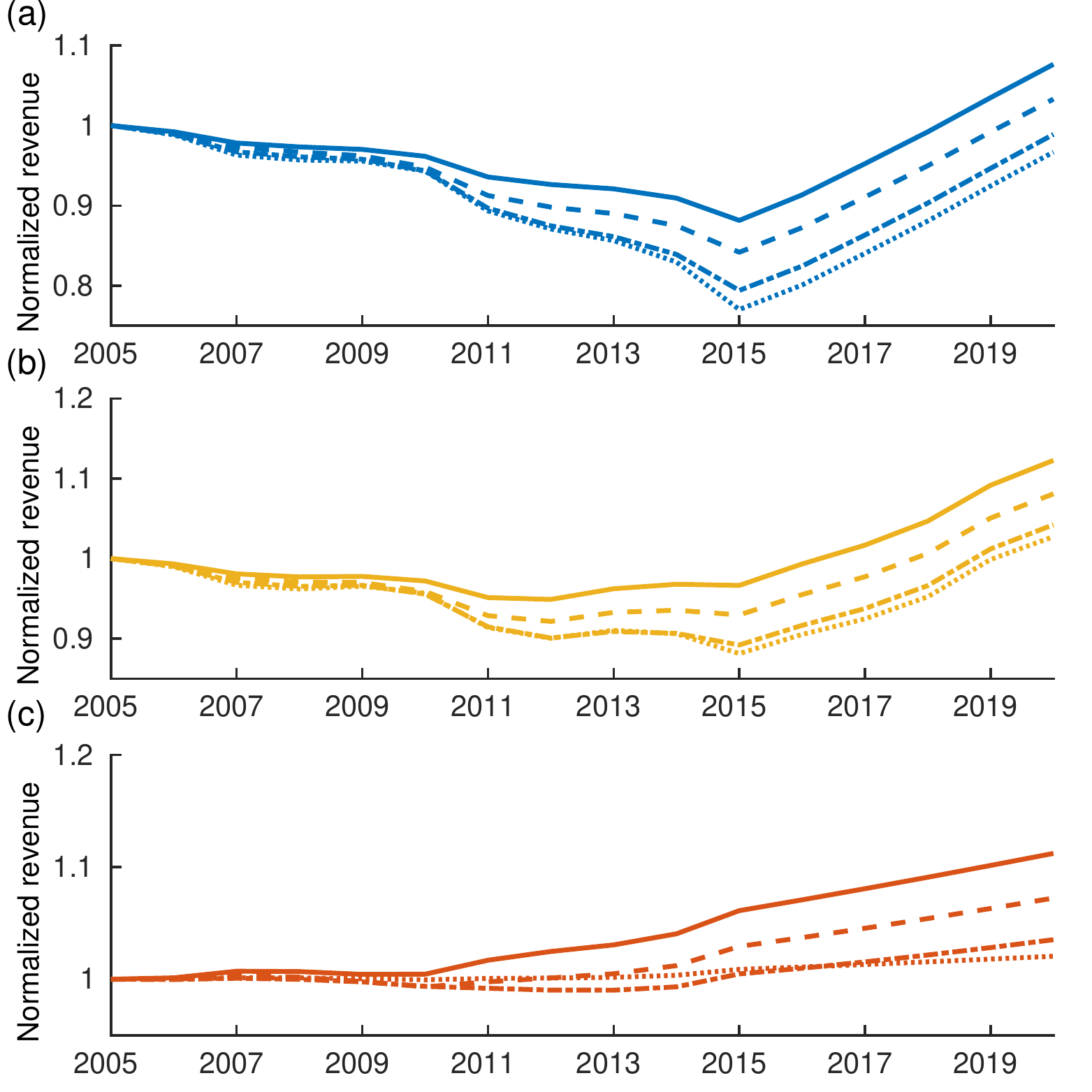}
\caption{Normalized revenues of super-peaking plants (functioning 1000 hours/year at maximal power; solid lines), peaking plants (2000 hours/year; dashed lines), 
load-following plants (5000 hours/year; dotted-dashed lines) and base-load plants (8000 hours/year; dotted lines) under 
the three scenarios "as long as possible" (panel a), "expected path" (panel b), "exact substitution" (panel c) of must-run withdrawal.}\label{flex}
\end{figure}

To understand better the trends discussed above, we finally investigate different types of productions characterized only by the number of hours $N_{\rm op}$
they operate at maximal power $P_{\mathrm{D}}^{\max}$ per year with no further constraint.
Accordingly, we consider four classes of power plants which are (i) super-peaking plants, functioning $N_{\rm op}=$ 1000 hours per year at peak power, 
(ii) peaking plants $N_{\rm op}=2000$, (iii) load-following plants $N_{\rm op}=5000$ and (iv) base-load plants $N_{\rm op}=8000$. We calculate their revenues using 
Eq.~\eqref{eq:gaindam}, with the constraint $\sum_k P_{\mathrm{D},k} = P_{\mathrm{D}}^{\max} \, N_{\rm op}$.
Fig.~\ref{flex} shows the evolution of these revenues as the energy transition unfolds in Germany, for our three
scenarios for must-run reduction. We see first, that regardless of must-run reduction, peak production plants always have higher revenues and second, that 
faster must-run withdrawal leads to smaller reductions in revenues. In particular, there is no significant decrease in revenue in the exact substitution scenario, 
for which thermal plants are retired in direct proportion to the penetration increase of new RES.

These results indicate that the currently very low electricity prices in Europe (and the poor revenues of the hydroelectricity sector)
are due to an overcapacity of electricity production more than anything else. Thermal plants
are currently retired too slowly compared to the rate at which the penetration of new RES increases. We therefore conjecture that 
the paradox of the energy transition can be overcome by retiring thermal plants faster.

\section{Conclusions}\label{section:concl}

Our interest in this manuscript has been to investigate the conditions under which the energy transition in the electric sector can proceed in Europe, without financially
jeopardizing flexible productions. This is a key issue, since increasing the penetration of new RES in the continental European grid will eventually
require sizeable power reserves that can be mobilized fast and often to compensate unavoidable, stochastic fluctuations in RES productions. Such reserves 
already exist (an example is hydroelectricity), however, they are currently under strong financial stress because of very low electricity prices. It is of paramount importance to figure out 
how long this stress will last and how it can be reduced in order to secure the future of flexible productions and with them, the harmonious
unfolding of the energy transition. This is what we called "overcoming the paradox of the energy transition".

To investigate this issue, we constructed a physico-economic model of the European grid. We observed a strong correlation between day-ahead electricity prices and 
residual loads all over Europe, and from that correlation, we constructed an electricity price solely based on the residual load, Eqs.~\eqref{eq:lin_correl} and \eqref{pda}. 
This price enabled us to investigate future revenues of various types of power plants. 
We found that three ingredients determine the occurrence, magnitude and duration of the paradox of the energy transition: (i) the rate at which the energy transition proceeds and RES penetration is increased, 
(ii) the mix of new RES and (iii) the rate of must-run withdrawal. In particular, if the must-run is kept high as RES penetration increases, electricity prices go down
with base prices well below the marginal cost of any flexible production. 
The hydroelectric sector in continental Europe is currently suffering from very low electricity prices, and our results indicate that
this situation is mostly due to surplus must-run capacity.
The duration of this paradoxical situation will decisively depend on how fast surplus thermal production capacities are 
withdrawn to compensate for the increased production from new RES. We advocate a faster withdrawal, for instance by substituting flexible 
gas-powered plants for coal-fired plants, which would additionally reduce greenhouse gas emissions faster.

The energy transition is accompanied by an increased need of production flexibility as it proceeds. Plants with large power will be needed more frequently,
and it is expectable that they will operate on a peak mode with 2000 hours of operation per year or less. It may well be that different business plans will
be developed for such plants, with different financial tools to reward not the energy produced, but the ancillary services provided. How such incentives will be introduced
remains speculative. 
Our results suggest that, even without them, peak and super-peak power plants should soon benefit again from improved financial conditions. 

As a final comment we note that different scenarios with different consumption curves can be investigated with the model presented above. 
In particular, active demand response can be incorporated into the model, as we discuss in \ref{appendix}.

\section*{Acknowledgment}

This work has been supported by the Swiss National Science Foundation. We thank T. Coletta, R. Delabays and M. Tyloo for their useful comments on the manuscript, M. Emery and M. Schmid for comments and Swissgrid data and R. Whitney for proofreading the manuscript. 
\appendix
\section{Incorporating active demand response}\label{appendix}
We sketch how active demand response (ADR) can be incorporated into our model. 
With ADR, the residual load is given by
\begin{equation}
R(t)=L(t)+\delta L(t)-P^{\rm PV}(t)-P^{WT}(t)-P^{\rm MR}=R_0(t)+\delta L(t)\,,
\end{equation}
where changes in the load profile due to ADR are included in $\delta L(t)$ and $R_0$ is the residual load without ADR, as in 
Eq.~\eqref{eq:rc1}. ADR can be deployed for various reasons, for instance to reduce electricity costs of end users or to mitigate 
load fluctuations on the distribution network. In both instances, ADR tries to reduce variations in the residual load, and we incorporate this
goal in an optimizing procedure which we briefly describe. For the sake of simplicity, we do not incorporate specific load constraints
such as comfort temperature intervals for ADR with thermostatically controlled loads.
The only constraint on the ADR profile  is 
\begin{equation}
\big|\delta L(t)\big|<\delta L^{\max}\,,\forall t\,,
\end{equation}
where $\delta L^{\max}$ is the maximal ADR power. We further assume that the annual consumption remains unchanged,
\begin{equation}
\intop_{t_i}^{t_f}\delta L(t){\rm d}t=0 \, .
\end{equation}
Recent estimates of the potential of ADR indicate that only a fraction of the total consumption can be shifted,
\begin{equation}
\intop_{t_i}^{t_f}\big|\delta L(t)\big|{\rm d}t\le2\sigma\intop_{t_i}^{t_f}L(t){\rm d}t\,,
\end{equation}
with $\sigma \simeq 0.01$ giving the maximal fraction of the total consumption that can been shifted, while the maximal ADR power
$\delta L^{\max}$ is about 10 \% of the maximal load $L^{\max}$ (roughly corresponding to 7 GW in 
Germany)~\citep{gils2016economic}. These numbers may seem rather small, however they have been obtained assuming 
a broad load participation in ADR~\citep{gils2016economic}.

Our procedure is to compute the ADR profile that minimizes the fluctuations of the residual load,
\begin{equation}
\underset{\delta L}{\min}\big[\mathrm{Var}(R)\big]=\underset{\delta L}{\min}\big[\mathrm{Var}(L^{+}-L^{-}+R_0)\big] \, ,
\end{equation}
where we defined $L^\pm(t)= {\rm max}[0,\pm \delta L(t)]$. We linearly increase $\sigma$ and $\delta L^{\max} /L^{\max}$
from 0 in 2015 to $\sigma=0.01$ and $\delta L^{\max} /L^{\max}=0.1$ in 2025. 
Because we neglect specific load constraints on ADR, our results likely overestimate the impact of ADR on the
residual load, and therefore on electricity prices. 

\begin{figure}[h!]
\includegraphics[width=\columnwidth]{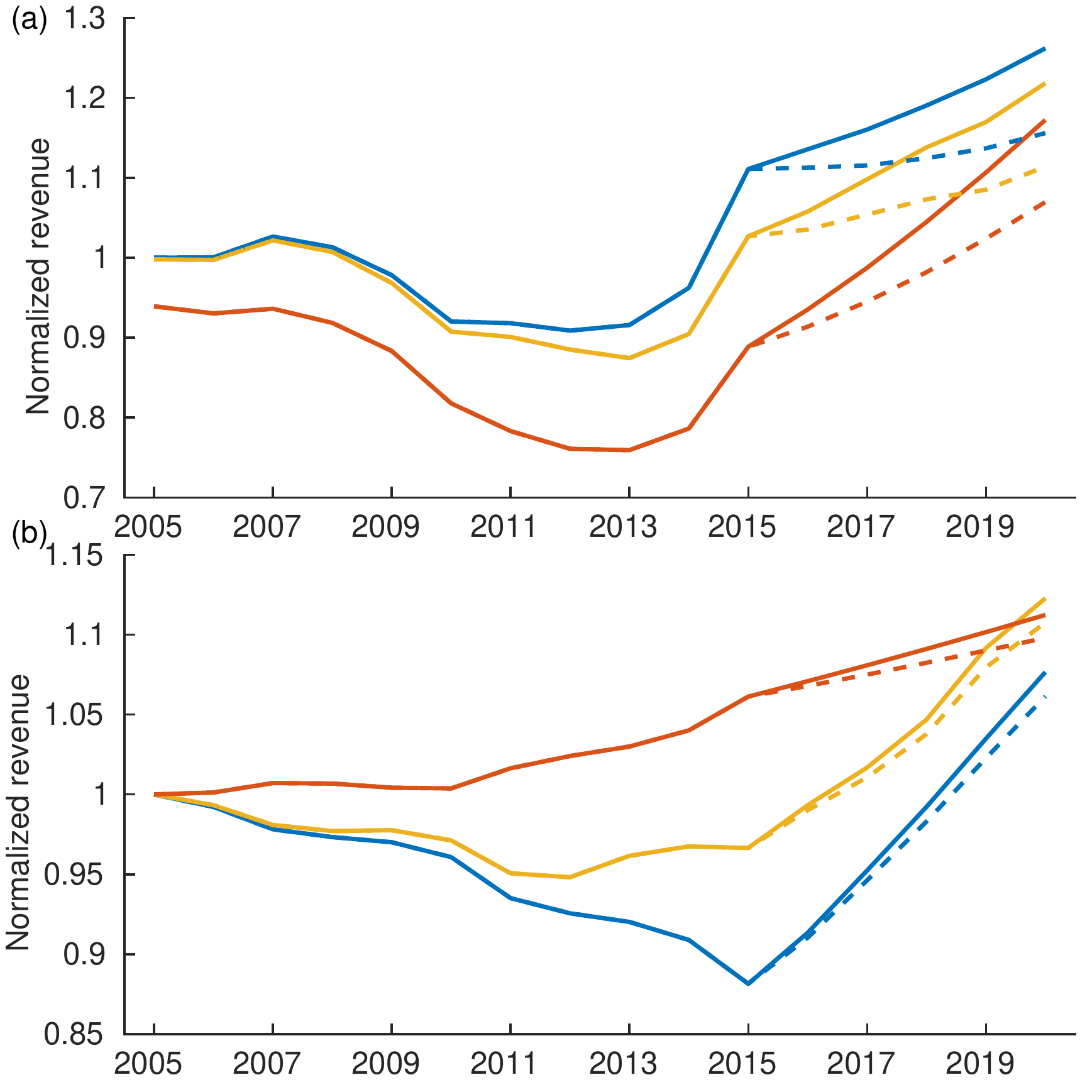}
\caption{(a) Revenues of a PS plant in Germany with (plain) and without (dashed) ADR. (b) Revenues of a conventional dam hydroelectric plant in Germany with (plain) and without (dashed) ADR. ADR linearly evolves from zero in 2015 to its maximal potential in 2025, with $\delta L^{\max}=7$[GW] and $\sigma=0.01$ \citep{gils2016economic}.}\label{DR}
\end{figure}

Fig.~\ref{DR} (a) shows the effect of ADR on the revenues of a PS plant. ADR being a form of storage, it competes with
PS and reduces its revenues, however, the effect is rather moderate, with 2020 revenues still exceeding those of 2005. 
Fig.~\ref{DR} (b) shows the revenues of a conventional dam hydroelectric plant, which are even less affected by ADR than those of 
the PS plant. 

Qualitatively, our conclusions stated in the main text remain valid.

\section*{References}
%\bibliographystyle{elsarticle-harv}
%%\bibliographystyle{elsarticle-num}
%%\biboptions{longnamesfirst,angle,semicolon}
%\bibliography{biblio}

\begin{thebibliography}{27}
\expandafter\ifx\csname natexlab\endcsname\relax\def\natexlab#1{#1}\fi
\expandafter\ifx\csname url\endcsname\relax
  \def\url#1{\texttt{#1}}\fi
\expandafter\ifx\csname urlprefix\endcsname\relax\def\urlprefix{URL }\fi

\bibitem[{Auer and Haas(2016)}]{auer2016integrating}
Auer, H., Haas, R., 2016. On integrating large shares of variable renewables
  into the electricity system. Energy 115, 1592--1601.

\bibitem[{Christensen et~al.(2012)Christensen, Hurn, and
  Lindsay}]{christensen2012forecasting}
Christensen, T.~M., Hurn, A.~S., Lindsay, K.~A., 2012. Forecasting spikes in
  electricity prices. International Journal of Forecasting 28, 400--411.

\bibitem[{Cl{\`{o}} et~al.(2015)Cl{\`{o}}, Cataldi, and Zoppoli}]{clo2015merit}
Cl{\`{o}}, S., Cataldi, A., Zoppoli, P., 2015. {The merit-order effect in the
  Italian power market: The impact of solar and wind generation on national
  wholesale electricity prices}. Energy Policy 77, 79--88.

\bibitem[{Cludius et~al.(2014)Cludius, Hermann, Matthes, and
  Graichen}]{cludius2014merit}
Cludius, J., Hermann, H., Matthes, F.~C., Graichen, V., 2014. {The merit order
  effect of wind and photovoltaic electricity generation in Germany 2008--2016:
  Estimation and distributional implications}. Energy Economics 44, 302--313.

\bibitem[{CREG(2015)}]{creg2015price}
CREG, 2015. {The price spikes observed on the Belgian day-ahead spot exchange
  Belpex on 22 September and 16 October 2015}. Tech. rep., Commission for
  electricity and gas regulation.

\bibitem[{Denholm and Hand(2011)}]{denholm2011grid}
Denholm, P., Hand, M., 2011. Grid flexibility and storage required to achieve
  very high penetration of variable renewable electricity. Energy Policy 39,
  1817--1830.

\bibitem[{ENTSO-E(2015{\natexlab{a}})}]{entsoe2015transparency}
ENTSO-E, 2015{\natexlab{a}}. {ENTSO-E Transparency platform}.
  \url{https://transparency.entsoe.eu/}.

\bibitem[{ENTSO-E(2015{\natexlab{b}})}]{entsoe2015tyndp}
ENTSO-E, 2015{\natexlab{b}}. {TYNDP 2016: Scenario Development Report}.

\bibitem[{{EPEX SPOT}(2015)}]{epex2015annual}
{EPEX SPOT}, 2015. Annual report.
  \url{https://www.epexspot.com/en/extras/download-center/activity_reports}.

\bibitem[{Guittet et~al.(2016)Guittet, Capezzali, Gaudard, Romerio, Vuille, and
  Avellan}]{guittet2016study}
Guittet, M., Capezzali, M., Gaudard, L., Romerio, F., Vuille, F., Avellan, F.,
  2016. Study of the drivers and asset management of pumped-storage power
  plants historical and geographical perspective. Energy 111, 560--579.

\bibitem[{Haas et~al.(2013)Haas, Lettner, Auer, and Neven}]{haas2013looming}
Haas, R., Lettner, G., Auer, H., Neven, D., 2013. The looming revolution: how
  photovoltaics will change electricity markets in europe fundamentally. Energy
  57, 38--43.

\bibitem[{Nicolosi(2010)}]{nicolosi2010wind}
Nicolosi, M., 2010. {Wind power integration and power system flexibility -- An
  empirical analysis of extreme events in Germany under the new negative price
  regime}. Energy Policy 38, 7257--7268.

\bibitem[{Nicolosi(2012)}]{nicolosi2012economics}
Nicolosi, M., 2012. The economics of renewable electricity market integration.
  an empirical and model-based analysis of regulatory frameworks and their
  impacts on the power market. Ph.D. thesis, Universit{\"a}t zu K{\"o}ln.

\bibitem[{NordPool(2015)}]{nordpool2015annual}
NordPool, 2015. Annual report.
  \url{http://www.nordpoolspot.com/About-us/Annual-report/}.

\bibitem[{OMIE(2015)}]{omie2015main}
OMIE, 2015. Main results of the electricity market.
  \url{http://www.omie.es/files/mercado_electrico_ing.diptico_web_pdf.pdf}.

\bibitem[{OTE(2015)}]{ote2015annual}
OTE, 2015. Annual report. \url{http://www.ote-cr.cz/about-ote/annual-reports}.

\bibitem[{Pagnier and Jacquod(2017)}]{pagnier2017predictive}
Pagnier, L., Jacquod, P., 2017. A predictive pan-european economic and
  production dispatch model for the energy transition in the electricity
  sector. In: PowerTech, 2017 IEEE Manchester. IEEE.

\bibitem[{Paraschiv et~al.(2014)Paraschiv, Erni, and
  Pietsch}]{paraschiv2014impact}
Paraschiv, F., Erni, D., Pietsch, R., 2014. {The impact of renewable energies
  on EEX day-ahead electricity prices}. Energy Policy 73, 196--210.

\bibitem[{Rehman et~al.(2015)Rehman, Al-Hadhrami, and Alam}]{rehman2015pumped}
Rehman, S., Al-Hadhrami, L.~M., Alam, M.~M., 2015. Pumped hydro energy storage
  system: A technological review. Renewable and Sustainable Energy Reviews 44,
  586--598.

\bibitem[{Saarinen et~al.(2015)Saarinen, Dahlb{\"a}ck, and
  Lundin}]{saarinen2015power}
Saarinen, L., Dahlb{\"a}ck, N., Lundin, U., 2015. Power system flexibility need
  induced by wind and solar power intermittency on time scales of 1--14 days.
  Renewable Energy 83, 339--344.

\bibitem[{Schill(2014)}]{schill2014residual}
Schill, W.-P., 2014. Residual load, renewable surplus generation and storage
  requirements in germany. Energy Policy 73, 65--79.

\bibitem[{Schlachtberger et~al.(2016)Schlachtberger, Becker, Schramm, and
  Greiner}]{schlachtberger2016backup}
Schlachtberger, D.~P., Becker, S., Schramm, S., Greiner, M., 2016. Backup
  flexibility classes in emerging large-scale renewable electricity systems.
  Energy Conversion and Management 125, 336--346.

\bibitem[{Schweiger et~al.(2017)Schweiger, Rantzer, Ericsson, and
  Leuenburg}]{schweiger2017potential}
Schweiger, G., Rantzer, J., Ericsson, K., Leuenburg, P., 2017. {The potential
  of power-to-heat in Swedish district heating systems}. Energy, in Press.

\bibitem[{Sensfuss et~al.(2008)Sensfuss, Ragwitz, and
  Genoese}]{sensfuss2008merit}
Sensfuss, F., Ragwitz, M., Genoese, M., 2008. {The merit-order effect: A
  detailed analysis of the price effect of renewable electricity generation on
  spot market prices in Germany}. Energy policy 36, 3086--3094.

\bibitem[{Swiss Federal Office of Energy(2016)}]{ofen2016weekstat}
Swiss Federal Office of Energy, 2016. Weekly report on water level of
  reservoirs.
  \url{http://www.bfe.admin.ch/themen/00526/00541/00542/00630/index.html?lang=en&dossier_id=00766}.

\bibitem[{Th{\"{u}}rler(2014)}]{thuerler2014statistik}
Th{\"{u}}rler, G., 2014. {Statistik der Wasserkraftanlagen der Schweiz}.
  \url{http://www.bfe.admin.ch/themen/00490/00491/index.html?lang=de&dossier_id=01049}.

\bibitem[{Ueckerdt et~al.(2015)Ueckerdt, Brecha, Luderer, Sullivan, Schmid,
  Bauer, B{\"o}ttger, and Pietzcker}]{ueckerdt2015representing}
Ueckerdt, F., Brecha, R., Luderer, G., Sullivan, P., Schmid, E., Bauer, N.,
  B{\"o}ttger, D., Pietzcker, R., 2015. Representing power sector variability
  and the integration of variable renewables in long-term energy-economy models
  using residual load duration curves. Energy 90, 1799--1814.

\bibitem[{von Roon and Huber(2010)}]{roon2010modeling}
von Roon, S., Huber, M., 2010. Modeling spot market pricing with the residual
  load. Tech. rep., {Forschungsstelle f\"ur Energiewirtschaft}.
  
  
\bibitem[{Gils(2016)}]{gils2016economic}
Gils, H.~C., 2016. Economic potential for future demand response in
  Germany--modeling approach and case study. Applied Energy 162, 401--415.

\end{thebibliography}

\end{document}